\documentclass[letterpaper]{article} 
\usepackage[]{aaai2026}  % DO NOT CHANGE THIS
\usepackage{times}  % DO NOT CHANGE THIS
\usepackage{helvet}  % DO NOT CHANGE THIS
\usepackage{courier}  % DO NOT CHANGE THIS
\usepackage[hyphens]{url}  % DO NOT CHANGE THIS
\usepackage{graphicx} % DO NOT CHANGE THIS
\usepackage{natbib}  % DO NOT CHANGE THIS AND DO NOT ADD ANY OPTIONS TO IT
\usepackage{caption} % DO NOT CHANGE THIS AND DO NOT ADD ANY OPTIONS TO IT
\usepackage{float} %设置图片浮动位置的宏包
\usepackage{subfigure} %插入多图时用子图显示的宏包
\usepackage{multirow}
\usepackage{threeparttable}
\usepackage[table,xcdraw]{xcolor}
\usepackage{algpseudocode}
\usepackage{makecell}
\usepackage{colortbl}
\usepackage{bm}
\usepackage{xspace}
\usepackage{amsmath}
\usepackage{float}
\usepackage{amsfonts}
\usepackage{booktabs}
\usepackage{threeparttable} 
\usepackage{array}
\usepackage{bigstrut}
\usepackage{adjustbox}
\usepackage{newfloat}
\usepackage{listings}
\usepackage{marvosym}
\DeclareCaptionStyle{ruled}{labelfont=normalfont,labelsep=colon,strut=off} % DO NOT CHANGE THIS
\floatstyle{ruled}
\newfloat{listing}{tb}{lst}{}
\floatname{listing}{Listing}
\title{DeepRAHT: Learning Predictive RAHT for Point Cloud Attribute Compression}
\author {
    Chunyang Fu\textsuperscript{\rm 1}, 
    Tai Qin\textsuperscript{\rm 2}, 
    Shiqi Wang\textsuperscript{\rm 1}\thanks{corresponding author},
    Zhu Li\textsuperscript{\rm 3}
}
\affiliations {
    \textsuperscript{\rm 1}City University of Hong Kong, Hong Kong SAR, China\\
    \textsuperscript{\rm 2}Peking University, Beijing, China\\
    \textsuperscript{\rm 3}University of Missouri-Kansas City, Kansas City, USA\\
    chunyang.fu@my.cityu.edu.hk, qintai@stu.pku.edu.cn, shiqwang@cityu.edu.hk, zhu.li@ieee.org
}

\usepackage{bibentry}
\newcommand{\ie}{{\emph{i.e.}}\xspace}
\newcommand{\eg}{{\emph{e.g.}}\xspace}
\newcommand{\etal}{{\emph{et al.}}\xspace}
\newcommand{\aka}{{\emph{aka}}\xspace}
\begin{document}

\maketitle

\begin{abstract}
    Regional Adaptive Hierarchical Transform (RAHT) is an effective point cloud attribute compression (PCAC) method. However, its application in deep learning lacks research. In this paper, we propose an end-to-end RAHT framework for lossy PCAC based on the sparse tensor, called DeepRAHT. The RAHT transform is performed within the learning reconstruction process, without requiring manual RAHT for pre-processing. We also introduce the predictive RAHT to reduce bitrates and design a learning-based prediction model to enhance performance. Moreover, we devise a bitrate proxy that applies run-length coding to entropy model, achieving seamless variable-rate coding and improving robustness. DeepRAHT is a reversible and distortion-controllable framework, ensuring its lower bound performance and offering significant application potential. The experiments demonstrate that DeepRAHT is a high-performance, faster, and more robust solution than the baseline methods.
\end{abstract}

\begin{links}
    \link{Project Page}{https://github.com/zb12138/DeepRAHT}
\end{links}
\section{Introduction}
A 3D point cloud is a set of points that consists of geometry (\ie, position of points) and attributes (\eg, color, normal vectors, and reflectance). As a common representation of 3D data, point clouds have been widely applied in various practical applications such as mixed reality, autonomous vehicles, and high-resolution mapping~\cite{usecase}. However, the massive points' unstructured and unordered nature presents challenges for existing compression techniques. The point cloud geometry and attribute compression can be performed jointly~\cite{alexiou2020towards,10980362}, but the dominant approaches in recent research~\cite{UnicornI, UnicornII, guo2024tsc, fang20223dac} and the MPEG geometry-based point cloud compression standard G-PCC~\cite {schwarz2018emerging} compress the geometry first and then compress the attributes based on the reconstructed geometry. The geometry compression of point clouds has made monumental progress in recent research~\cite{fu2022octattention,songEHEM, UnicornI,wang2025unipcgc,you2025reno}, particularly in the various deep learning-based approaches. However, the attribute compression has not been sufficiently explored. This paper studies the lossy compression of point cloud attributes, assuming that the geometry has been reconstructed.

Point cloud attributes are essential and have various applications. For instance, colors provide the most basic visual information, reflectance is used for object detection in LiDAR, and normals are utilized for surface reconstruction and rendering. Furthermore, the recently popular Gaussian splatting data~\cite{3dgs} is also related to point clouds, as the parameters of the Gaussian primitives can be viewed as the attributes. One important method for lossy point cloud attribute compression is the Region-Adaptive Hierarchical Transform (RAHT)~\cite{RAHT}, which has been adopted as the core method in the G-PCC standard~\cite{gpcc}, demonstrating excellent complexity and performance. However, only a few research about deep learning studies on RAHT. 3DAC~\cite{fang20223dac} is the first framework that uses a deep entropy model for coding RAHT transform coefficients, but it merely uses the manually crafted RAHT (\ie, same as the original implementation) to generate transform coefficients first. Then a deep entropy model was devised to learn the distribution of coefficients. Since the manual RAHT is non-differentiable, 3DAC can only optimize the bitrate. 3DAC also ignored the predictive RAHT, a crucial improvement in the current G-PCC RAHT standard~\cite{gpcc}. Prediction can significantly reduce the uncertainty of transform coefficients, and thus, encoding the residuals, not the coefficients, can remarkably reduce the bitrate. However, the learning of the prediction of RAHT has been largely overlooked. The lack of end-to-end framework limits the application of RAHT in deep learning, such as joint learning with RAHT and other components (\eg, deblocking task). 

In this paper, we propose the first end-to-end differentiable RAHT framework called \textbf{DeepRAHT}. We also integrated a learnable prediction model. In our method, the multi-scale RAHT transform is performed during the learning reconstruction process, without requiring a manual RAHT in advance. The end-to-end design provides high application potential for our framework, allowing joint optimization of RAHT and other components. DeepRAHT has an identical framework to the MPEG G-PCC and builds the performance lower bound on it. Furthermore, DeepRAHT is entirely reversible and distortion controllable. Our main contributions are as follows:

\begin{itemize}
\item We implement the first end-to-end differentiable RAHT framework, which has an identical structure to the G-PCC reference software (\ie, tmc13v14, \aka G-PCCv14) and establishes a performance lower bound on it.

\item We develop a learnable prediction model incorporating grandfather scale context. This model validates the learnability of DeepRAHT and achieves over 24\% BD-rate gains compared to G-PCCv14.

\item We propose a rate proxy to utilize run-length coding as the entropy coder, replacing the entropy bottleneck~\cite{balle2018variational} used in most works. Run-length coding is more robust for our framework and achieves seamless variable-rate coding.
\end{itemize}

\section{Related Work}
\subsection{Traditional PCAC Methods}
The key aspect of point cloud attribute compression (PCAC) is to explore attribute correlations through geometry structures. Mainstream PCAC methods primarily utilize prediction, projection, and transformation techniques. The prediction tree \cite{waschbusch2004progressive} is the earliest related method in prediction-based methods. G-PCC~\cite{gpcc} predicting transform branch also performs predictions in refinement levels divided by levels of detail. Projection-based methods, such as MPEG V-PCC~\cite{graziosi2020overview}, convert 3D point sets into 2D images, enabling the application of existing image and video compression standards (e.g., JPEG and H.265)~\cite{mekuria2016design,li2020efficient}. Transformation-based methods focus on compressing attributes in the frequency domain, where energy is more concentrated. Graph Fourier Transform (GFT) and its variants~\cite{cohen2016attribute,shao2017attribute,xu2020predictive,10013714,10445988} have proven effective but are too complex when performing eigen decomposition. Queiroz~\etal,~\cite{RAHT} propose an efficient adaptive Haar wavelet transform called Region-adaptive Hierarchical Transformation (RAHT), which avoids the eigen decomposition in GFTs and is adopted in the MPEG G-PCC standard. Our framework achieves the parallelization of RAHT, resulting in the low complexity of DeepRAHT.

\subsection{Learning-based PCAC Methods}
Alexiou~\etal~\cite{alexiou2020towards} pioneered compressing point cloud attributes using dense 3D convolution. Following this, Deep-PCAC\cite{Deep-PCAC} introduced another approach based on point convolution. FoldingNet~\cite{quach2020folding} projected point clouds into 2D images, applying image codecs for compression. LVAC~\cite{isik2022lvac} utilized Implicit Neural Representation (INR) for point cloud attribute compression. Fang~\etal~\cite{fang20223dac} proposed 3DAC, the first learning-based method to encode RAHT coefficients directly; however, the RAHT used was non-differentiable and lacked the prediction component. Concurrently, SparsePCAC\cite{sparsePCAC} made progress by leveraging stacked sparse convolutional layers. Despite these innovations, their lossy compression performance still lagged the early G-PCC test model (\ie, G-PCCv14).

Recently, some methods have made significant progress and claim to outperform G-PCCv14. 3CAC~\cite{3CAC} and CNet~\cite{CNET} were the early attempts for lossless PCAC. TSC-PCAC~\cite{guo2024tsc} enhanced SparsePCAC by incorporating Transformer-based modules for inter-channel context regression. Li~\etal~\cite{10301698} introduced the first network based on graph dictionary learning. Mao~\etal proposed SPAC\cite{SPAC} for lossy PCAC, a sampling-based framework utilizing residual networks and multiple slices. Recently, Unicorn~\cite{UnicornII} stands as the state-of-the-art deep learning framework for point cloud compression, employing average pooling to obtain multiple scales and sparse convolutions to code attribute residuals. These methods demonstrated superior results across most datasets, although specific cases revealed some robust issues.

\section{Methodology}

\subsection{Overview}

Given a 3D voxelized point cloud with $N$ points denoted by $P=(p,a)$, $p\in \mathbb{N}^{N \times 3}$ is the set of points' positions and $a\in \mathbb{R}^{N}$ is the corresponding attribute\footnote{A single channel of attributes is discussed here, which is more general and the other channels are similar.}. Suppose the input point cloud is $P_0$, we aim to lossily compress $a_0$ with known $p_0$. In DeepRAHT, $P_0$ is first pooled $s$ times by sum-pooling with stride $2\times2\times2$, and thus generates $s$ scales $\{P_1,...,P_m,...,P_s\}$. Define $A_m$ as the sum of attributes associated with points encompassed by the node in $P_m$, thus from the definition of sum-pooling,
\begin{equation}
 A_{m+1,i}=\sum_{j\in \mathcal{N}_{i}}A_{m,j}, \ \ m=0,1,...,s-1,
\label{Eq1}
\end{equation}
where $i$ is the parent node in $P_{m+1}$ with at most eight child nodes ${j}$ in $P_{m}$ after pooling. $A_0 = a_0$ is the input attributes.

\begin{figure}[]
    \centering
    \includegraphics[scale=0.77]{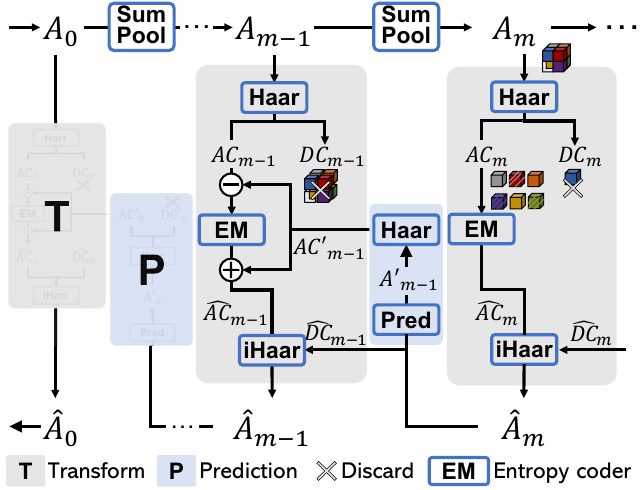}
    \caption{Overview of DeepRAHT. $A_0$ is the input attributes and $\hat{A}_0$ is the reconstructed attributes.
    }
    \label{overview}
\end{figure}

The overall framework of DeepRAHT is illustrated in the Fig. \ref{overview}. Initially, sum-pooling is performed $s$ times to generate multi-scale point clouds. The encoding process starts from the last scale $s$ in a top-down manner. Point cloud at each scale is applied with the $\textbf{T}$ransform model, while the $\textbf{P}$rediction model is optional (\eg, prediction is disabled in scale $m$ for illustration). In the $\textbf{T}$ model, $A_m$ is decomposed into the alternating coefficient $AC_m$ and the direct coefficient $DC_m$ using the \textbf{Haar} module. The $AC_m$ or its residual is encoded and decoded by the entropy coder $\textbf{EM}$, while $DC_m$ (excluding the root $DC_s$) is equivalent to $A_{m+1}$ and has been encoded at a higher scale and thus discarded. 

Meanwhile, reconstruction also proceeded while encoding. The \textbf{iHaar} module performs an inverse Haar transform to reconstruct $\hat{A}_{m}$ using the decoded $\widehat{AC}_m$ and the reconstructed $\widehat{DC}_m$, where $\widehat{DC}_{m}$ is reconstructed from the higher scale $\hat{A}_{m+1}$. The reconstructed $\hat{A}_{m}$ is subsequently used to reconstruct $\widehat{DC}_{m-1}$ similarly. In the $\textbf{P}$ model, $\hat{A}_{m}$ also predicts $AC'_{m-1}$ for the residual coding of $AC_{m-1}$. The processes $\mathbf{T}$ and $\mathbf{P}$ are repeated until encoding $A_0$.

The decoding process also starts from scale $s$ and is consistent with the reconstruction process, and $\widehat{A}_0$ represents the decoded point cloud attributes. Table~\ref{notation} is the notations.

\begin{table}[t]
    \centering
    \begin{tabular}{c|c}
        \toprule
        \textbf{Notation}                 & \textbf{Description}           \\ \midrule
        $A$                               & sum of attributes              \\
        $w$                               & number of points in node       \\
        $a=A/w$                           & averaged attributes            \\
        $g=A/\sqrt{w}$ & normalized attributes \\
        $g_\text{L},...,g_\text{HHH}$ & transform coefficients         \\ \bottomrule
    \end{tabular}
    \caption{Notations.}
    \label{notation}
\end{table}

\subsection{Transform Model}
This section demonstrates the \textbf{Haar} and \textbf{iHaar} modules and their convolution-based implementation. Region Adaptive Hierarchical Transform (RAHT) is a variation of the Haar wavelet transform, and it was first proposed in ~\cite{RAHT}. In this paper, we use a dyadic version of RAHT~\cite{dyadicRAHT}, which improves decorrelation of the high-frequency coefficients in original RAHT. 

\begin{figure}[t]
    \centering
    \includegraphics[scale=0.45]{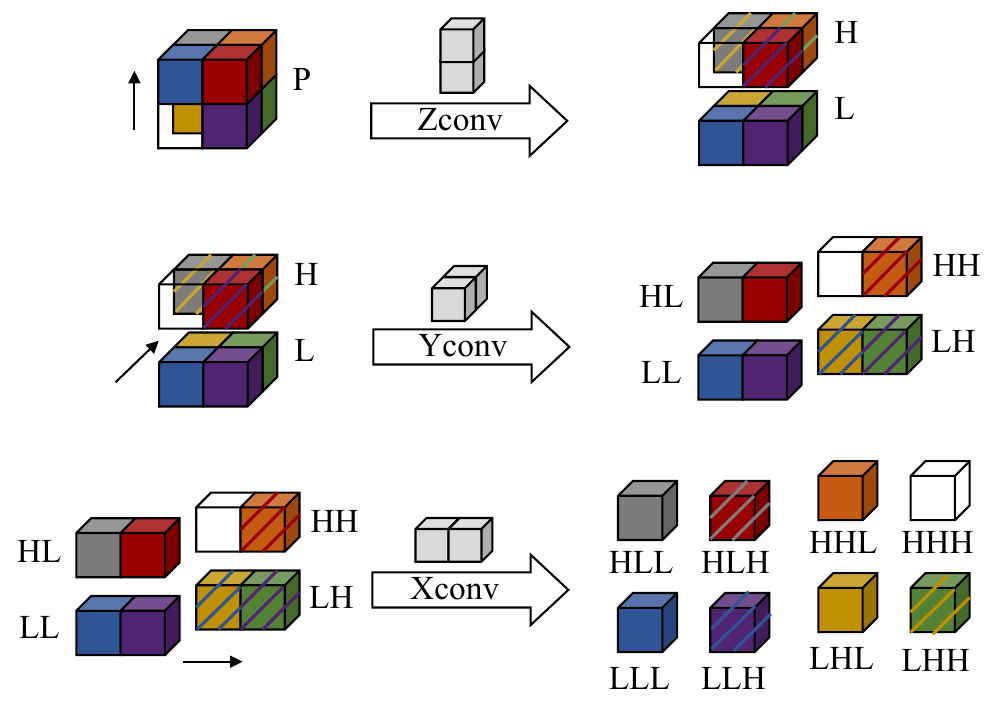}
    \caption{
        Dyadic RAHT decomposition.
    }
    \label{DyadicRAHT}
\end{figure}

For each $2\times2\times2$ voxel, the Haar transform decomposes eight nodes into one direct coefficient and seven alternating coefficients. Fig.~\ref{DyadicRAHT} shows an example, where nodes are with indexes $j$ at scale $m$. Note that one of them is transparent, indicating that it encompasses no points. Denote $g_{{m}}$ as the normalized attributes as the node features for the input, 
\begin{equation}
 g_{{m},j} = A_{{m},j}/\sqrt{w_{{m},j}},\ \ j\in \mathcal{N}_i.
\label{eq2}
\end{equation}
$w_{{m},j}$ is the number of points of $p_0$ encompassed by node. $A_{{m},j}$ is the sum of attributes of those points, and $i$ is the index of output DC (node $\text{LLL}$) in $P_{m+1}$, defined in Eqn.~\eqref{Eq1}. 

Z-axis decomposition is applied first and illustrated as an example. The normalized attributes $g_1, g_2\in\{g_{{m},j}\}$ of the lower and upper node pairs (\eg, purple and red) are transformed according to the following formula,
\begin{equation}
    \left[\begin{array}{l}
            g_\text{L} \\
            g_\text{H}
        \end{array}\right]=\frac{1}{\sqrt{w_{1}+w_{2}}}\left[\begin{array}{cc}
            \sqrt{w_{1}}  & \sqrt{w_{2}} \\
            -\sqrt{w_{2}} & \sqrt{w_{1}}
        \end{array}\right]\left[\begin{array}{l}
            g_{1} \\
            g_{2}
        \end{array}\right],
    \label{eq3}
\end{equation}
where weight $w_1$ is the number of the points encompassed by the lower node (\eg, purple node) and $w_2$ is the counterpart of the upper node (\eg, red node).  $g_\text{L}$ and $g_\text{H}$ are the output coefficients for the low frequency and high frequency coefficients (\ie, shaded) nodes, and they will serve as the normalized attributes for the following decompositions. The weights of the output nodes are assigned as $w_\text{L}=w_\text{H}=w_1+w_2$. Note that if one of the nodes in the pair is empty, the normalized attributes of the nonempty node will transfer directly to the low frequency, to be forced to undergo further decomposition. The counterpart of high frequency is set to zero: $g_\text{L}=g_{\text{nonempty}}, g_\text{H}=0, w_\text{H}=0$. After this decomposition, the input coefficients are converted to half low-frequency and half high-frequency coefficients. Similarly, two Y-axis decompositions for the \text{L}, \text{H} nodes, and four X-axis decompositions for the \text{LL}, \text{LH}, \text{HL}, \text{HH} nodes, are applied sequentially according to Eqn.~\eqref{eq3}.

After the decompositions of the three axes, $g_{m}$ are transformed into eight groups of coefficients $g_\text{LLL}, g_\text{LLH},..., g_\text{HHH}$. Note that their quantities are not equal due to the empty nodes. $g_\text{LLL}$ group constitutes the direct coefficient (DC), and it can be proved that DC is equivalent to the normalized attributes of the following scale (see supplementary for the proof),
\begin{equation}
    DC_{m,{\mathcal{N}_i}}\equiv g_{\text{LLL}_{m,{\mathcal{N}_i}}} = A_{{m+1},i}/\sqrt{w_{{m+1},i}} = g_{m+1,i},
    \label{Eq4}
\end{equation}
where $w_{{m+1},i}=\sum_{j\in\mathcal{N}_i} w_{{m},j}$ and $A_{{m+1},i}$ is defined in Eqn.~\eqref{Eq1}. In this regard, in terms of DC, the Haar transform is equivalent to the sum-pooling with a stride of $2\times2\times2$ described in Eqn.~\eqref{Eq1}, and followed by a square-root normalization. From this equation, $DC_{m}$ is equivalent to $A_{{m+1}}$, thus $DC_{m}$ is discarded at encoding, because $A_{{m+1}}$ has been encoded in the higher scale. Specifically, the root DC in the final scale (\ie, $ DC_s$) is encoded directly. The other seven coefficients constitute alternating coefficients (AC) and will be compressed using the entropy coder.

In the reconstruction, from Eqn.~\eqref{Eq4}, DC can be reconstructed from the decoded attributes from the higher scale
\begin{equation}
 \widehat{DC}_{m} = \widehat{A}_{{m+1}}/\sqrt{w_{{m+1}}},
\label{eq5}
\end{equation}
and AC are decoded from the bitstream by the entropy model. The \textbf{iHaar} module operates in the reverse manner and finally obtains reconstructed attributes $\widehat{A}_{{m}}$.

\subsubsection{Implementation}
G-PCC~\cite{gpcc} implements the above RAHT by a manual framework based on C++, and it is not differentiable. We use Minkowski Sparse Tensor~\cite{sparseConv} to build the differentiable dyadic RAHT. A sparse tensor $\{C,F\}$ has coordinate field $C\in\mathbb{N}_0^{N \times 3}$ and associated features $F\in\mathbb{R}^{N \times k}$. Thus, the sparse tensor $\bm{A}_0$\footnote{Bold font represents sparse tensor and italic font denotes its features $A=\bm{A}. F$.} for $A_0$ can be defined as $\{p_0,a_0\}$. Then the formula Eqn.~\eqref{Eq1} is equivalent to
\begin{equation}
 \bm{A}_{m+1}=\operatorname{SumPooling}(\bm{A}_{m},k=2^3,s=2^3),
\label{Eq6}
\end{equation}
where ${k}$ means the kernel size and ${s}$ means the stride. The $\operatorname{SumPooling}$ can also be used to calculate the number of points encompassed by nodes. Define the initial weight sparse tensor $\bm{w}_0=\{p_0,\bm{1}\}$, then
\begin{equation}
 \bm{w}_{m+1}=\operatorname{SumPooling}(\bm{w}_{m},k=2^3,s=2^3).
\label{w}
\end{equation}

\begin{figure}[t]
    \centering
    \includegraphics[scale=1]{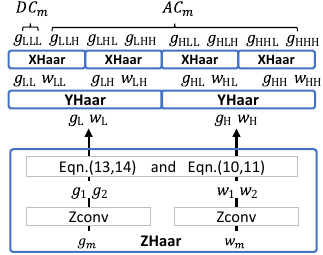}
    \caption{
        Details of \textbf{Haar} module.
    }
    \label{HaarM}
\end{figure}

To model Haar decompositions, we use sparse convolution to model each axis decomposition. Given Z-axis as an example, the details are shown in the \textbf{ZHaar} of Fig.~\ref{HaarM}, where $\operatorname{Zconv}$ is a sparse convolution and it is defined as, 
\begin{equation}
 \operatorname{Zconv}\equiv\operatorname{Conv}(i=1,o=2,k= s=(1, 1, 2)),
\end{equation}
where ${i}$ means input channels, ${o}$ means output channels. The initial weights for the convolution kernel are predefined as the identity matrix $\bm{I}_2$. In this way, nodes $g_1$ and $g_2$ are obtained and mapped to the two output channels. The definitions of $\operatorname{Yconv}$ and $\operatorname{Xconv}$ are similar, but the kernel size and stride are $(1, 2, 1)$ and $(2, 1, 1)$, respectively.

Denote $\text{f}\in\{\text{P}, \text{L},\text{H},\text{LL},..,\text{HH},\text{LLL},...\text{HHH}\}$ (omit the default scale $m$) as the frequencies in Fig.~\ref{DyadicRAHT}, and denote ${w}_\text{f}$ as the weight (points number) for each node. ${w}_\text{L}\dots,{w}_\text{HH}$ can be drawn by,
\begin{align}
    \bm{w}_{1},\bm{w}_{2} & = \operatorname{HaarConv}(\bm{w}_\text{f}) \label{w1w2} \\
    \bm{w}_{\text{fL}}    & = \bm{w}_{1}+\bm{w}_{2}                                 \\
    \bm{w}_{\text{fH}}    & = \left\{
    \begin{array}{ll}
        \bm{0}             & \text{if } w_{1}=0 \text{ or } w_{2}=0 \\
        \bm{w}_{\text{fL}} & \text{otherwise},
    \end{array}
    \right.
\end{align}
where $\operatorname{HaarConv}\in\{\operatorname{Zconv},\operatorname{Yconv},\operatorname{Xconv}\}$ according to Fig.~\ref{DyadicRAHT}. Note that $\bm{w}_\text{P}=\bm{w}_{m}$ as the initial weights. $\text{fL}$ and $\text{fH}$ mean corresponding low and high frequency of the base frequency (\eg, if $\text{f}=\text{P}$, $\text{fH}=\text{H}$; if $\text{f}=\text{H}$, $\text{fL}=\text{HL}$).

Similarly, Eqn.~\eqref{w1w2} can be used to calculate the transform coefficients $\bm{g}_\text{f}$. Therefore, the formula Eqn.~\eqref{eq3} for forward Haar can be implemented by,
\begin{align}
    \bm{g}_1,\bm{g}_2 & =\operatorname{HaarConv}(\bm{g}_\text{f})        \\
    \bm{g}_\text{fL}  & = \alpha*\bm{g}_1 + \beta*\bm{g}_2  \label{eq13} \\
    \bm{g}_\text{fH}  & = \left\{
    \begin{array}{ll}
        \bm{0} \quad\quad\quad\text{if } w_{1}=0 \text{ or } w_{2}=0               \\
        -\beta*\bm{g}_1 + \alpha*\bm{g}_2 \quad\text{otherwise}, \label{eq14} \\
    \end{array}\right.
\end{align}
where,
\begin{align}
    \alpha & = \sqrt{{w_1}/(w_1+w_2)}, \beta=\sqrt{{w_2}/(w_1+w_2)}.
    \label{ab}
\end{align}
Note that $\bm{g}_\text{P}=\bm{g}_{m}$. The details of \textbf{Haar} module is shown in Fig.~\ref{HaarM}. Eqn.~\eqref{w1w2}$\sim$\eqref{ab} are repeated as $\text{f}$ from $\text{P}$ to $\text{HH}$ until $DC_{m} =g_\text{LLL}$ and $AC_{m} =\{g_\text{LLH},...,g_\text{HHH}\}$ are obtained.

The \textbf{iHaar} module employs $\operatorname{ConvolutionTranspose}$ to model the inverse of $\operatorname{HaarConv}$, \ie,
\begin{equation}
    \operatorname{iZconv} \equiv\operatorname{ConvT}(i=2, o=1, k=s=(1, 1, 2)).
\end{equation}
The implementation for the backward Haar is,
\begin{align}
    \bm{g}_1        & =\alpha*\bm{g}_\text{fL} - \beta*\bm{g}_\text{fH} \\
    \bm{g}_2        & =\beta*\bm{g}_\text{fL} + \alpha*\bm{g}_\text{fH} \\
    \bm{g}_\text{f} & =\operatorname{iHaarConv}([\bm{g}_1,\bm{g}_2]),
\end{align}
where $\alpha,\beta$ are defined in Eqn.~\eqref{ab}, $\operatorname{iHaarConv}\in\{\operatorname{iXconv},\operatorname{iYconv},\operatorname{iZconv}\}$, and $\text{f}$ is from $\text{HH}$ to $\text{P}$. Finally, attributes sum is reconstructed by $\widehat{A}_{m}=g_\text{P}*\sqrt{w_{m}}$.

\subsection{Prediction Model}
\label{Predictive RAHT}
In DeepRAHT, we encode from the coarsest layer $P_s$ to the finest layer $P_0$, and thus we can incorporate the prediction model. G-PCC version 14 introduces predictive RAHT in~\cite{predictiveRaht}, which up-samples the attributes of the current node by parent neighbors based on the inverse distance weighted prediction. In current G-PCC version 23, Wang \etal~\cite{predictiveRaht2} introduce utilizing the decoded sibling neighbors to enhance prediction accuracy. However, this auto-regressive technique will significantly increase decoding time if applied to batch processing-based methods in deep learning. Therefore, we only use the parent scales to design the prediction model as in G-PCCv14. We first introduce a learning implementation of IDW and then propose a prediction compensation module to enhance the prediction performance. Experiment results indicate the learning-based $\textbf{Pred}$ model can achieve an even better performance than that of sibling-based prediction.
\subsubsection{IDW Prediction}
% Below is the implementation of the IDW prediction module. We introduce an interpolation method to up-sample the parent point cloud. From Eqn.~\eqref{Eq6}, after $m$ times pooling, the coordinate of $\bm{A}_m$ becomes $C_m= 2^m\lfloor C_0/2^m \rfloor\ $. To obtain the same resolution as the point cloud $\bm{A}_{m-1}$, the points of $\bm{A}_m$ are directly copied and interpolated,
% \begin{equation}
% \begin{aligned}
% \operatorname{Interp}(\bm{A}_m) = \{C_m+2^{m-1}u, F_m\} \cup \bm{A}_m \\ 
% \text{for } u \in \{(0,0,1), (0,1,0), (0,1,1), \ldots, (1,1,1)\}
% \end{aligned}
% \end{equation}
% After the up-sampling, 
We design a sparse convolution for the inverse distance weighted (IDW) prediction. To ensure stability, IDW is performed in the averaged attributes domain. Suppose  $\bm{a}'_{m-1} = \operatorname{IDW}(\widehat{\bm{a}}_{m})$ is the predicted averaged attributes of scale $m-1$, and it is implemented as,
\begin{equation}
    \begin{aligned}
        \operatorname{IDW}(\widehat{\bm{a}}_m) \equiv \operatorname{Conv}(\operatorname{Unpool}(\widehat{\bm{a}}_m),k=3^3,s=1^3)
        \label{IDw}
    \end{aligned}
\end{equation}
where $\widehat{\bm{a}}_m\equiv \widehat{\bm{A}}_m/w_m$ is the averaged attributes, and $\widehat{\bm{A}}_m$ is the reconstructed attributes from parent scale,
$\operatorname{Unpool}$ is an unpooling operation with stride 2. The $\operatorname{Conv}$ is a normalized convolution where the weights of the convolution kernel sum to 1 and the proportions are predefined based on the distance from the kernel center. Specifically, the proportions of weights for center, faces, edges, and angles in kernel are set to 4, 3, 2, and 1.

\subsubsection{Prediction Compensation}
\begin{figure}[t]
    \centering
    \includegraphics[scale=0.8]{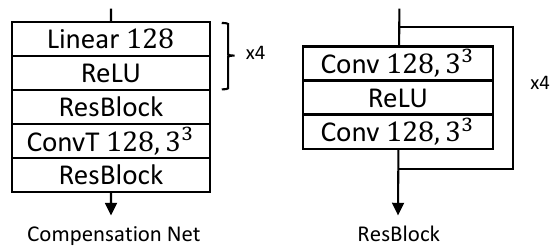}
    \caption{
        Prediction compensation module.
    }
    \label{compensation}
\end{figure}

As mentioned earlier, we do not utilize the sibling nodes to facilitate prediction. Instead, we propose a compensation module to enhance the prediction by leveraging context from the grandparent (\ie, $m+1$) scale. This approach avoids the auto-regression problem, thereby not increasing the decoding time too much. We use the prediction error between the grandfather and the father scales to compensate for the current prediction. Combining with Eqn.~\eqref{IDw}, the prediction after compensation is,
\begin{equation}
 \bm{a}'_{m-1}=\operatorname{Comp}(\widehat{\bm{a}}_{m}-\operatorname{IDW}(\widehat{\bm{a}}_{m+1})) + \operatorname{IDW}(\widehat{\bm{a}}_{m}),
\end{equation}
where $\operatorname{Comp}$ is the compensation module, which is shown in the Fig.~\ref{compensation} (omit the linear header of input and output). The prediction compensation module is designed empirically and composed of multiple stacked linear layers and convolutions with hidden layer dimension 128, kernel size $3^3$, including a transposed convolution with a stride of 2. 

The predicted averaged attributes $\bm{a}'_{m-1}$ yields predicted sum of attributes $\bm{A'}_{m-1}=\bm{a}'_{m-1}w_{m-1}$, and Haar transform is also applied to predict the AC coefficients $AC'_{m-1}=\textbf{Haar}(\bm{A'}_{m-1})$ as shown in the $\textbf{P}$ model. Finally, the residuals of AC,
\begin{equation}
 r_{m-1} = AC_{m-1} - AC'_{m-1}
\label{ori_res}
\end{equation}
are encoded.

Note that the prediction compensation can be removed based on the prediction performance, which can be evaluated on mean square error, with a cost of $s$ bits to signal to the decoder. This mechanism improves the robustness of DeepRAHT and ensures that the performance lower bound is G-PCCv14. The removal hardly occurred in the experiments, which indicated that the prediction compensation module was always practical.

\subsection{Entropy Coder}
Most deep learning-based compression methods utilize bottleneck~\cite{balle2018variational} in their entropy models. However, it is greatly affected by the variance of the data. To address this issue, we use the zero run-length coding~\cite{Run_length} (see supplementary for the details) as the entropy coder. Zero run-length coding is efficient for compressing data where there are large numbers of zeros. The coefficient residuals of RAHT are highly concentrated around zeros and thus zero run-length coding performs better, converges faster, and is more robust. Since the run-length coding is not differentiable, we propose a rate proxy based on the Laplace distribution to estimate its probabilistic model,
\begin{equation}
    q(r) = \int_{r-0.5}^{r+0.5}\mathcal{L}_{\mu, \sigma}(r)\, dr,
    \label{EM}
\end{equation}
where the mean $\mu$ and standard variance $\sigma$ are determined by experiments. The actual bitrate of run-length coding is approximately equal to the cross-entropy $R\approx\alpha H(p, q)$, where $p$ is the distribution of quantized residuals $\operatorname{Q}(r/qs)$, and $\alpha$ represents the proportion by which the run-length coding outperforms the proxy with the assumed Laplace distribution. $\operatorname{Q}$ is a $\operatorname{STE\_ROUND}$~\cite{STE}, and $qs$ is the quantization step. Note that the Haar transform in DeepRAHT is entirely reversible and thus the distortion comes from the quantization of the residuals, which only depends on the quantization step.

\subsection{Learning}
Since \textbf{Haar}, \textbf{iHaar}, \textbf{Pred} as well as the \textbf{EM} modules are end-to-end differentiable, the final reconstruction error can directly be obtained by $\ell_{recon}=\|a_0-\widehat{a}_0\|_2^2$. In the prediction model, the prediction loss $\ell_{pred}=\sum_m\|(a_m-a'_m)\|_2^2$ can also facilitate convergence. On the other hand, the loss of bitrate can be drawn from the rate proxy $\ell_{bits}=-\sum_m\log_2q(r_m/qs)$, where $q(r)$ is defined in Eqn.~\eqref{EM}. Thus, the total loss is,
\begin{equation}
\ell= \ell_{bits} + \lambda(\ell_{recon} + \ell_{pred}).
\end{equation}
$\lambda$ is the weight to balance the distortion and bitrate loss.

\begin{figure}[!h]
    \centering

    \begin{adjustbox}{max width=\textwidth} 
        \hspace{-0.018\textwidth}
        \subfigure{
            \includegraphics[scale=0.337]{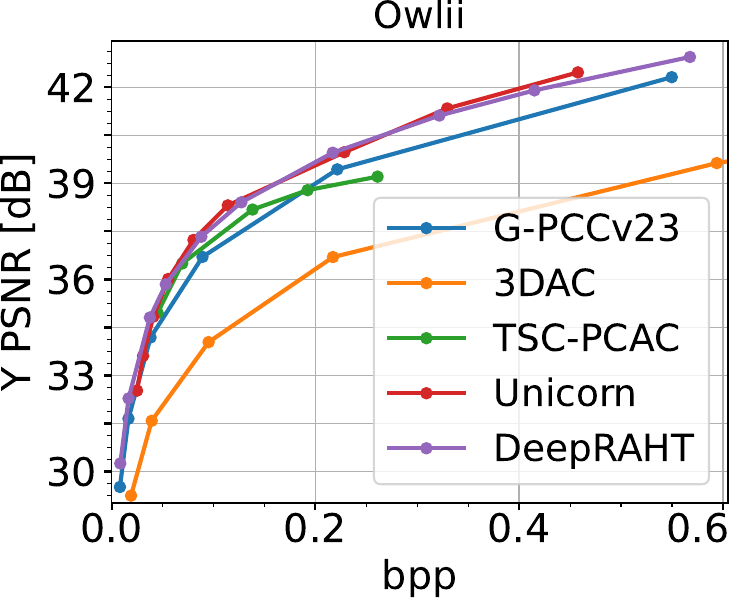}
        }
        \hspace{-0.012\textwidth}
        \subfigure{
            \includegraphics[scale=0.337]{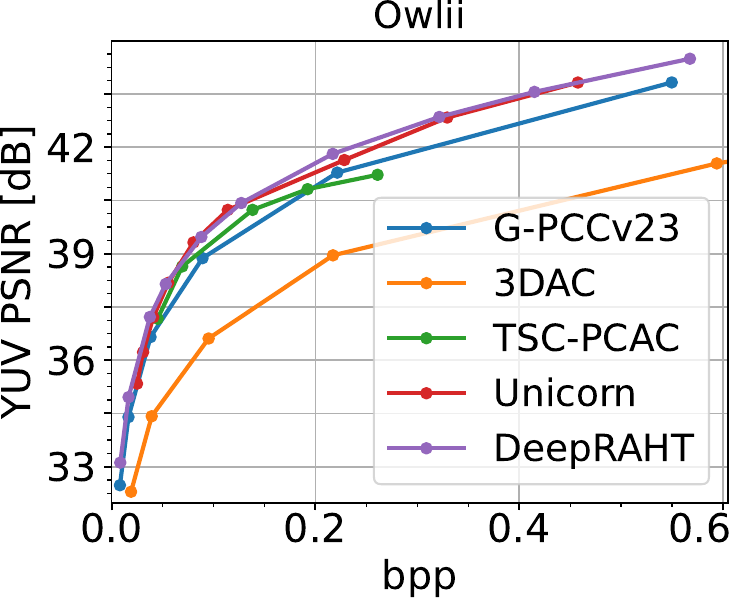}
        }
    \end{adjustbox}
    \begin{adjustbox}{max width=\textwidth} 
        \hspace{-0.018\textwidth}
        \subfigure{
            \includegraphics[scale=0.335]{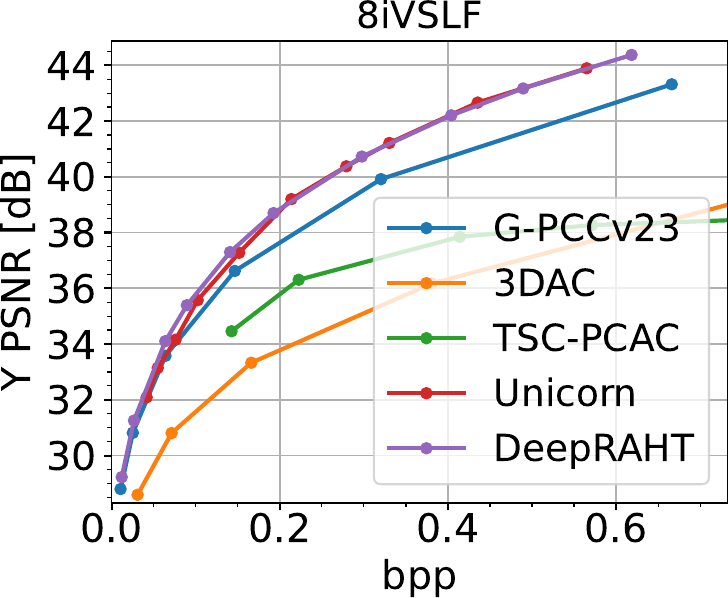}
        }
        \hspace{-0.012\textwidth}
        \subfigure{
            \includegraphics[scale=0.335]{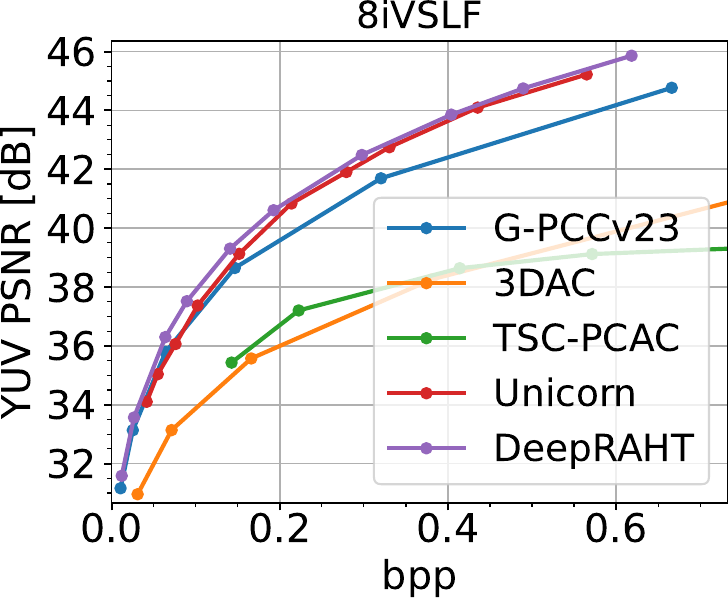}
        }
    \end{adjustbox}

    \begin{adjustbox}{max width=\textwidth} 
        \hspace{-0.018\textwidth}
        \subfigure{
            \includegraphics[scale=0.335]{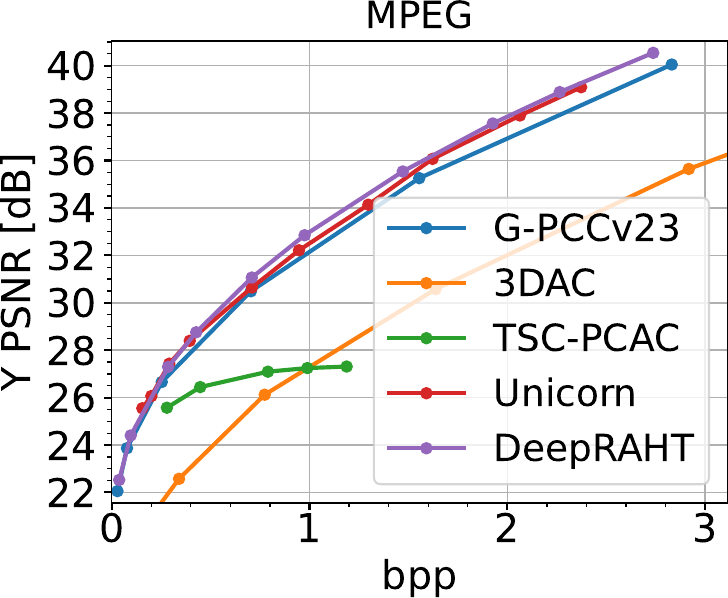}
        }
        \hspace{-0.012\textwidth}
        \subfigure{
            \includegraphics[scale=0.335]{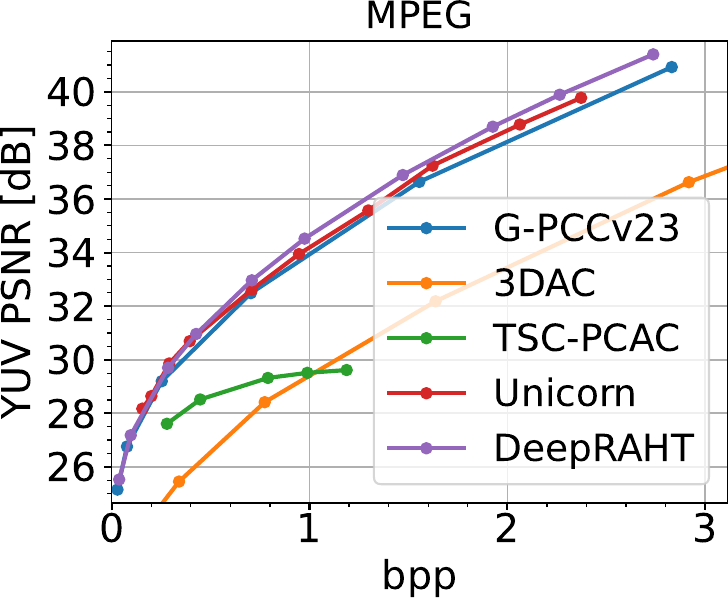}
        }
    \end{adjustbox}
\caption[]{R-D curves averaged over the datasets.}
\label{RDcurve}
\end{figure}

% \begin{figure}[!h]
%     \centering
%     \begin{adjustbox}{max width=\textwidth} 
%         \hspace{-0.015\textwidth}
%         \subfigure{
%             \includegraphics[scale=0.335]{Result/_0Fig/_8iVSLF/U/8iVSLF.pdf}
%         }
%         \hspace{-0.02\textwidth}
%         \subfigure{
%             \includegraphics[scale=0.335]{Result/_0Fig/_8iVSLF/V/8iVSLF.pdf}
%         }
%     \end{adjustbox}
% \caption[]{Chroma R-D curves averaged over 8iVSLF.}
% \label{RDcurveUV}
% \end{figure}

\section{Experiments}

\subsection{Experimental Setup}
\subsubsection{Datasets} We train our model on the RWTT~\cite{RAHT} dataset suggested in MPEG \cite{AIPCC}, which contains 568 real-world objects. The evaluation is performed on the first frames of Owlii~\cite{Owlii}, 8iVSLF~\cite{8i}, and MPEG CTC Samples~\cite{CTC}, which are popular testing sets in most of the PCAC works. 

\subsubsection{Evaluation Metrics} The peak signal-to-noise ratio (PSNR) is employed to assess reconstruction quality, and bits per point (bpp) is used to evaluate the compression ratio. The Bjøntegaard Delta Bit Rate (BD-BR) is used to assess the overall bitrate saving under the same quality.

\subsubsection{Baselines} \textbf{G-PCC}~\cite{gpcc} is the most widely used standard provided by MPEG. The latest public version (tmc13v23) on the Octree-RAHT branch is used for comparison. \textbf{3DAC}~\cite{fang20223dac} is the first learning-based method to code the manual RAHT coefficients. \textbf{TSC-PCAC}~\cite{guo2024tsc} is an auto-encoder method based on the Transformers and inter-channel context regression. \textbf{Unicorn}~\cite{UnicornII} is currently the state-of-the-art deep learning-based point cloud compression framework. All the learning-based methods are trained on RWTT.

\subsubsection{Implementation Details}
The transform and prediction are implemented by Pytorch, and the run-length coding is written by C++. Point cloud colors are compressed in YUV space. The training and testing of all methods are performed with a Core i9-13th CPU and one NVIDIA 4090 GPU (24 GB memory). The total scale $s$ is set to the geometry precision of the point cloud, and prediction starts from $s-2$ to $0$. We set $qs=8, \lambda=1/255$, batch sizes to 1, and use Adam optimizer with a learning rate of $0.0001$ for training. The validation losses details are in the supplementary.

\subsection{Experiment Results}

\begin{table}[!t]
    \centering
    \begin{threeparttable}
        \setlength{\tabcolsep}{0.55mm}
        {
            \centering
            \small
            \begin{tabular}{lrrrrrrrr}%{lcccccccc}
                \toprule
                \multirow{2}{*}{\textbf{Anchor}} & \multicolumn{2}{c}{\textbf{G-PCCv23}} & \multicolumn{2}{c}{\textbf{3DAC}} & \multicolumn{2}{c}{\textbf{TSC-PCAC}} & \multicolumn{2}{c}{\textbf{Unicorn}}                           \\
                     & \phantom{X}Y\phantom{X}                       & YUV                   & \phantom{X}Y\phantom{X}                  & YUV    & \phantom{X}Y\phantom{X}  & YUV & \phantom{X}Y\phantom{X}      & YUV   \\
                \midrule
                basketball           & -23.9&-24.8       & -69.1&-68.7           & -26.4&-31.0         & -7.8&-11.0\\
                dancer           & -24.4&-25.5       & -72.2&-71.4           & -19.2&-26.6         & -5.8&-10.9\\
                exercise           & -14.3&-16.8       & -64.9&-64.5             & -0.4&-2.7        & -10.0&-13.5\\
                model           & -13.0&-13.0       & -62.6&-61.9               & 0.2&9.2            & 8.0&7.1\\
   \textbf{Owlii AVG.}           & -18.9&-20.0       & -67.2&-66.6           & -11.4&-12.8          & -3.9&-7.1\\
    \midrule
            Thaidancer           & -13.2&-12.3       & -62.9&-63.6           & -34.8&-47.4            & 8.0&3.0\\
            boxer           & -23.0&-25.9       & -76.4&-75.9           & -82.5&-89.8        & -20.5&-25.7\\
            longdress           & -17.3&-16.5       & -67.7&-68.3           & -51.8&-61.7           & 1.1&-6.6\\
            loot           & -16.6&-20.0       & -73.0&-73.0           & -66.0&-73.1         & -6.9&-12.0\\
            redandblack           & -14.5&-13.3       & -76.5&-76.6           & -62.6&-72.3         & -8.6&-20.3\\
            soldier           & -16.1&-17.2       & -68.7&-67.7           & -46.5&-66.6          & -0.8&-3.8\\
\textbf{8iVSLF AVG.}           & -16.8&-17.5       & -70.9&-70.9           & -57.4&-68.5         & -4.6&-10.9\\
          \midrule
          Egyptian             & -8.6&-8.1       & -43.0&-44.0           & -83.8&-91.6            & NA&NA\\
            Facade           & -16.1&-15.3       & -76.9&-78.1           & -63.2&-63.6            & 8.0&7.1\\
             House             & -4.1&-3.8       & -79.1&-81.7           & -74.9&-85.4          & -6.9&-6.5\\
             Shiva             & -2.3&-1.5       & -41.6&-44.0           & -50.9&-58.6        & -10.4&-10.6\\
               ULB           & -15.3&-14.7           & NA&NA           & -69.6&-76.3            & NA&NA\\
             queen           & -29.4&-29.6       & -64.9&-65.8           & -70.7&-79.6            & NA&NA\\
             Staue             & -9.1&-7.9       & -60.8&-62.5           & -50.9&-57.1          & -4.5&-5.9\\
\textbf{MPEG AVG.}           & -12.1&-11.6       & -61.1&-62.7           & -66.3&-73.2          & -3.4&-4.0\\
\midrule
\textbf{Average}           & -15.9&-16.4       & -66.4&-66.7           & -45.0&-51.5          & -4.0&-7.3\\
                \bottomrule
            \end{tabular}}
        \begin{tablenotes}
            \footnotesize
            \item[] NA: \textit{The method fails to compress the data.}
        \end{tablenotes}
    \end{threeparttable}
    \caption{BD-BR gain (\%) against with baseline methods.}
    \label{BDBR}
\end{table}

\begin{figure}[b]
    \subfigure[GT]{
        \includegraphics[width=0.115\textwidth]{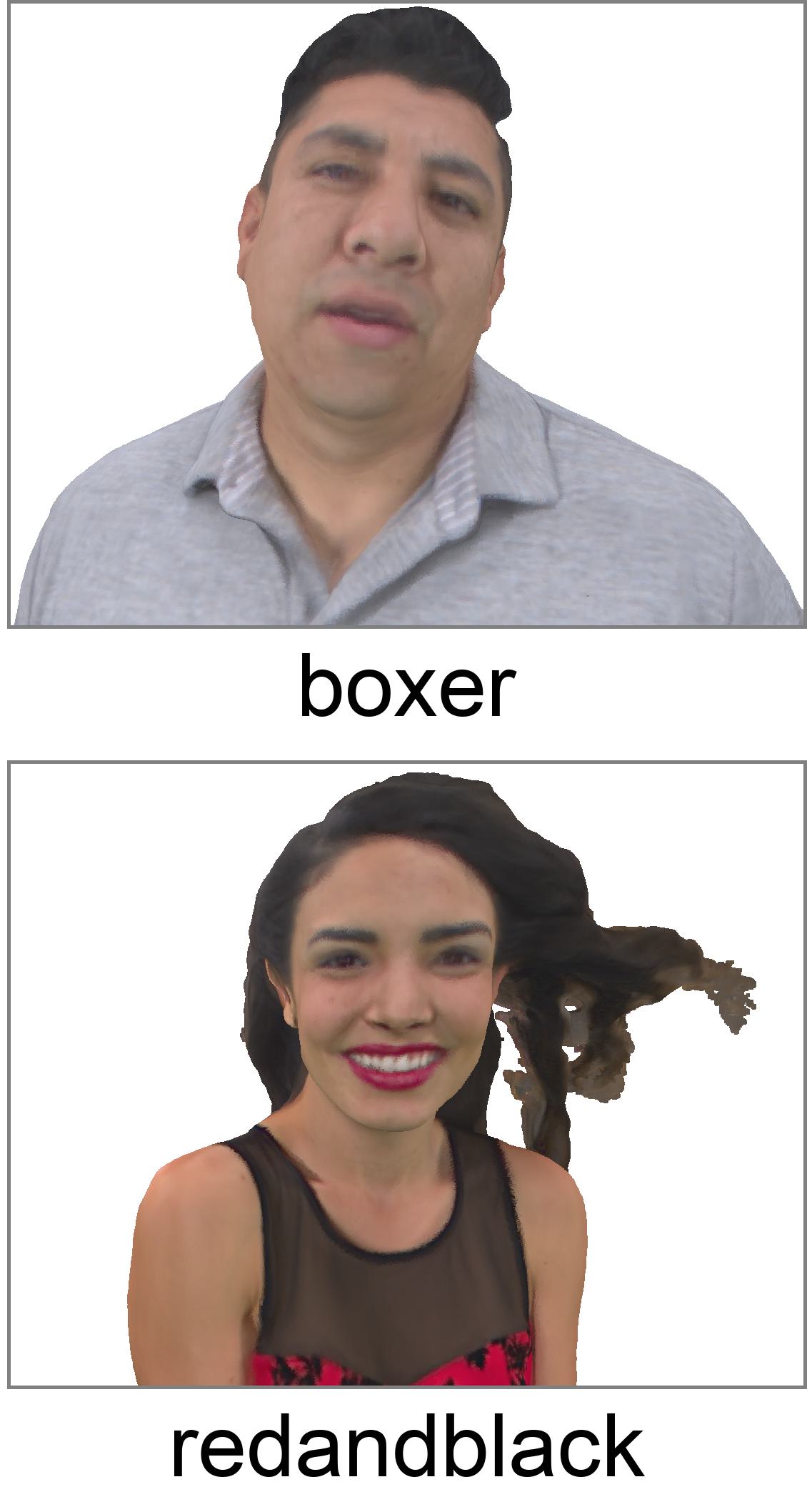}
    }
    \hspace{-0.018\textwidth} 
    \subfigure[G-PCC]{
        \includegraphics[width=0.115\textwidth]{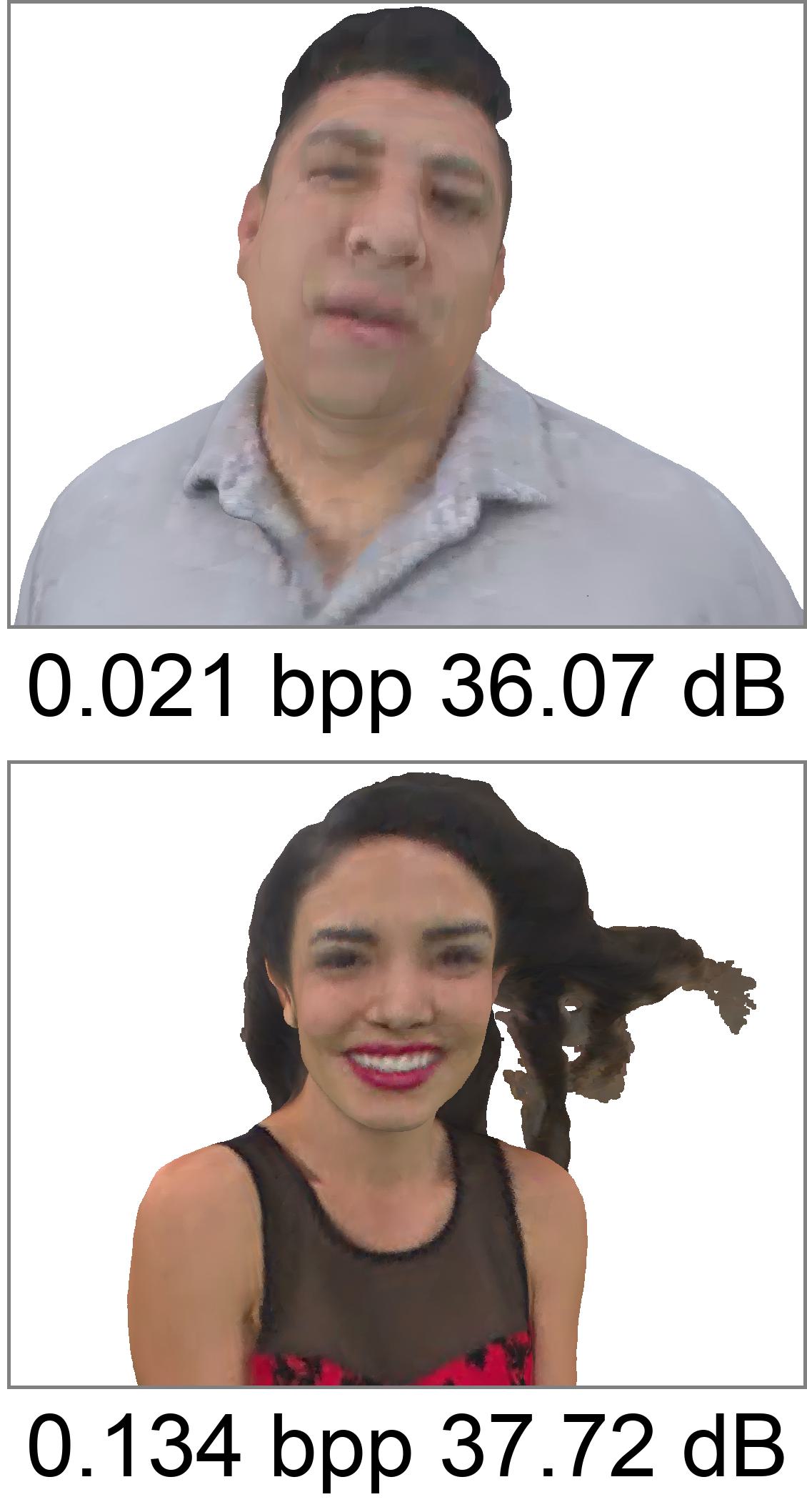}
    }
    \hspace{-0.018\textwidth} 
    \subfigure[Unicorn]{
        \includegraphics[width=0.115\textwidth]{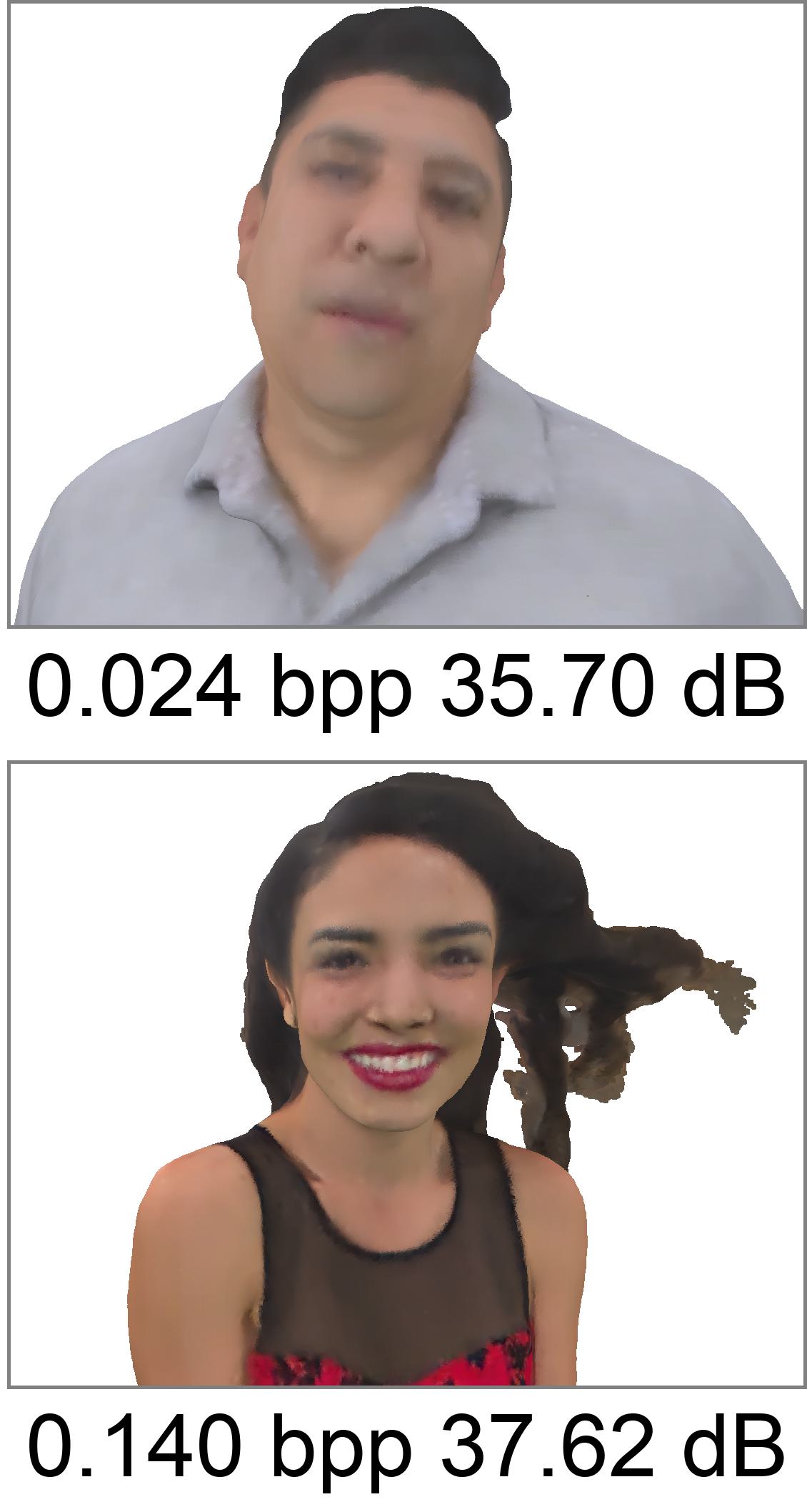}
    }
    \hspace{-0.018\textwidth}
    \subfigure[DeepRAHT]{
        \includegraphics[width=0.115\textwidth]{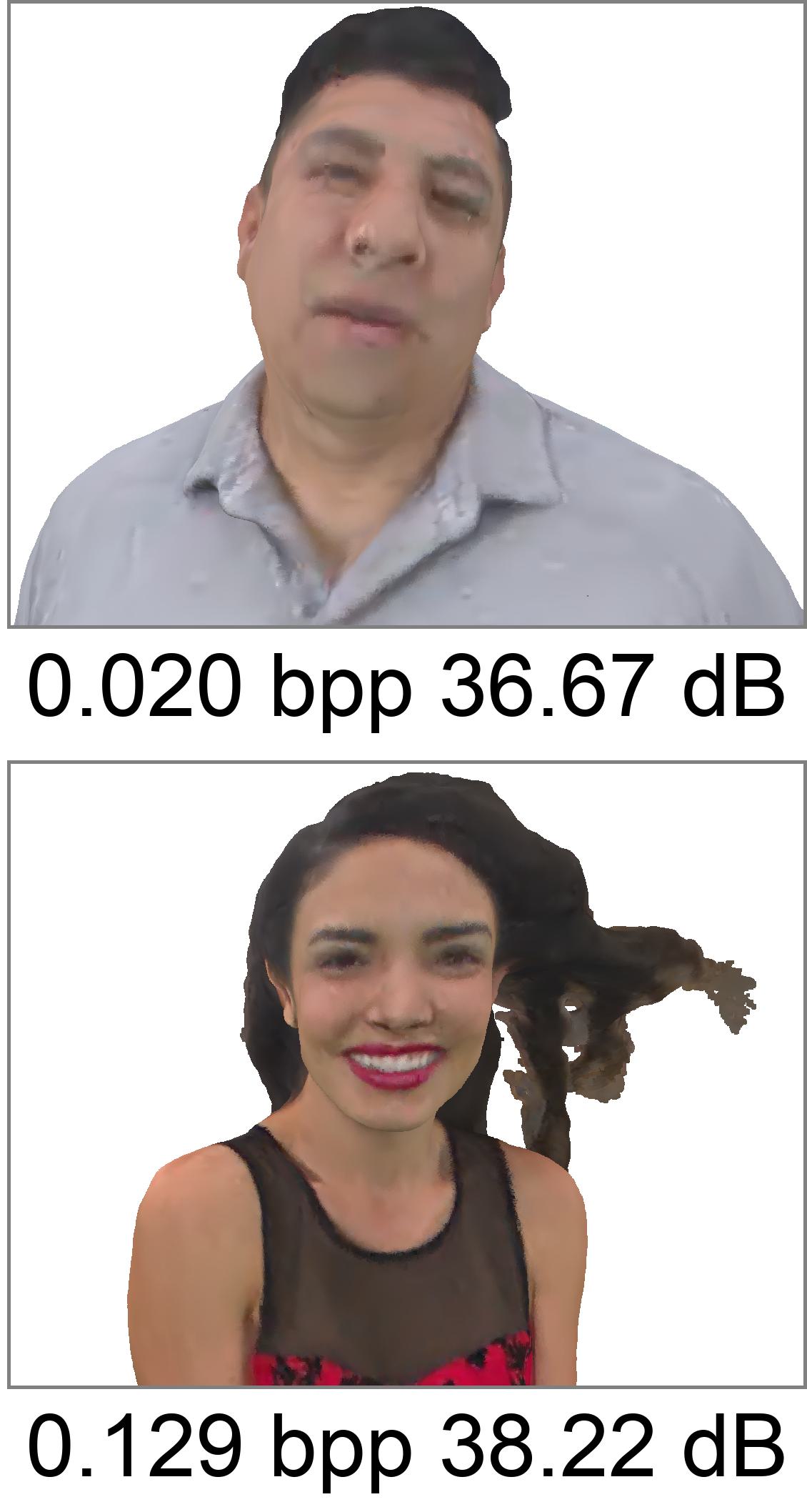}
    }
    \caption{Ground Truth (GT) and reconstructions of G-PCC, Unicorn and DeepRAHT. Bpp and Y PSNR are provided.}
    \label{vis}
\end{figure}

\begin{table}[b]
    \centering
    \setlength{\tabcolsep}{1.2mm}
    {
        \centering
        \small
        \begin{tabular}{cccccc}  
            \toprule
            \textbf{Method} & \textbf{3DAC} & \textbf{TSC-PCAC} & \textbf{Unicorn} & \textbf{DeepRAHT} \\
            \midrule
            Enc. Time       &  38.45 s            & 7.86 s            & 20.86 s          & \textbf{6.03 s}            \\ %2.0238,38.45,7.8568,20.86,7.77
            Dec. Time       &  51.71 s            & 26.87 s           & 14.99 s          & \textbf{5.74 s}            \\  % 1.75375,51.708,26.87,14.98,7.679
            \midrule
            Model Size      & \textbf{1 MB$\times$5} & 148 MB$\times$5   & 65 MB$\times$3   & 88 MB$\times$1    \\
            GPU Mem.        & 10 GB              & 22 GB             & 16 GB            & \textbf{8 GB}              \\
            \bottomrule
        \end{tabular}}
        \caption{Complexity comparison tested on 8iVSLF.}
    \label{Complexity}
\end{table}

\subsubsection{Quantitative BD-BR Gains}
The quantitative comparison of DeepRAHT with baseline methods is presented in Table~\ref{BDBR}. DeepRAHT outperforms all baseline methods, achieving a 16.4\% improvement over G-PCCv23 and a 7.3\% YUV BD-BR improvement over Unicorn on average. The averaged R-D curves are shown in Fig.~\ref{RDcurve}. DeepRAHT outperforms all the baseline methods across a wide bitrate range. Meanwhile, DeepRAHT demonstrated a significant improvement in chrominance, achieving a BD-BR gain of 20.5\% on U and 20.8\% on V compared to Unicorn. Detailed R-D curves are provided in the supplementary. Notably, DeepRAHT is more robust and successfully compressed and surpasses G-PCC across all data, while other learning-based methods struggled with certain large or sparse point clouds.

\subsubsection{Qualitative Visualization}
The visualization comparison is shown in Fig.~\ref{vis}. It can be observed that DeepRAHT significantly reduces artifacts compared to G-PCC. Unicorn produces smoother reconstructions but irreversibly loses details, such as the texture of the boxer's shirt and facial details of the redandblack, while DeepRAHT preserves more details. This advantage comes from the fact that DeepRAHT is reversible, where the distortion only depends on the quantization. More comparisons are provided in the supplementary.

\begin{figure}[t]
    \centering
    \includegraphics[scale=0.55]{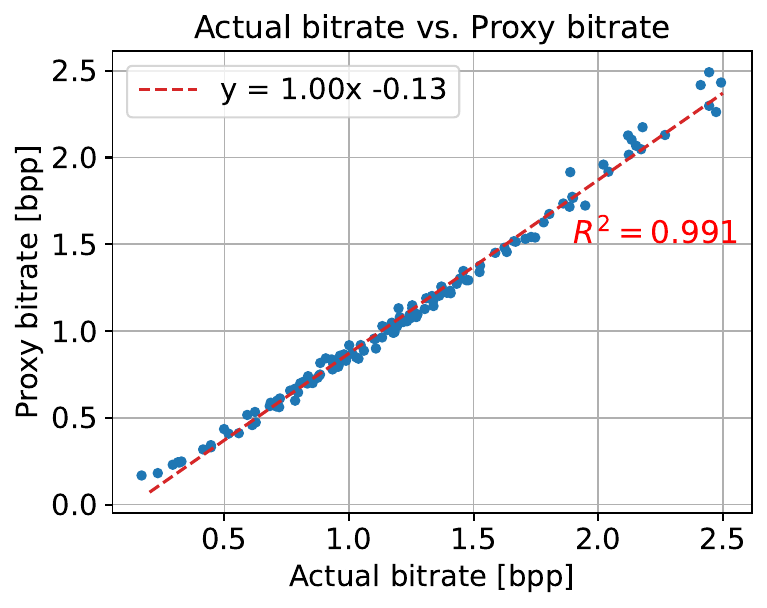}
    \caption{
        Bitrate of run-length coding and the proxy.
    }
    \label{rate}
\end{figure}

\subsubsection{Rate Proxy of Run-length Coding}
We propose a rate proxy within the entropy model to predict the rate of run-length coding. We obtain the parameters $\alpha=0.425,\mu=0,\sigma=0.2$ in Eqn.~\eqref{EM} by fitting the actual data. Fig.~\ref{rate} from the testing data illustrates the actual bitrate alongside the predicted values. The coefficient of determination is 0.991, demonstrating the accuracy of the proposed rate proxy. 

\subsubsection{Variable-rate Coding}
Variable-rate coding requires a single neural model to achieve lossy compression at varying rates and qualities. In contrast, 3DAC and TSC require training multiple models for variable rates because the entropy model they use heavily depends on the variance of the data. Unicorn aims to address this challenge by introducing the Adjustable Quantization Layer; however, a single model can only accommodate approximately three bitrates. We propose run-length coding to resolve this issue, demonstrating strong robustness to data with varying variance under a Laplace distribution. As a result, the rate points in Fig.~\ref{RDcurve} are obtained by adjusting $qs=\{8, 10, 12, 16, 24, 32, 48, 64, 128, 224\}$ on one highest rate checkpoint without needing extra training.

\subsubsection{Complexity}
The complexity comparison was evaluated and averaged across all bitrates and frames on the 8iVSLF dataset, which contains an average of 3251505 points per point cloud. The test point clouds are divided into blocks within $2*10^6$ points by KD-tree to avoid memory overflow. As shown in Table~\ref{Complexity}, DeepRAHT is faster than all the baseline methods in both encoding and decoding. The GPU memory usage of DeepRAHT is only 8 GB, suggesting that it can be much faster if it increases the number of points in the blocks. Notably, our model only needs a single checkpoint with 88 MB, maintains a moderate model size.  This efficiency is attributed to our relatively shallow network architecture and the efficient sparse convolution, which provides a significant advantage for practical applications.

% [{'ply': 'Data/8iVSLF/Static/loot_viewdep_vox12.ply', 'RAHT vs RAHT+Pred BR': -48.212974985740374}]
% [{'ply': 'Data/8iVSLF/Static/loot_viewdep_vox12.ply', 'RAHT vs RAHT+Pred+Comp BR': -59.20944668905916}]
% [{'ply': 'Data/8iVSLF/Static/loot_viewdep_vox12.ply', 'G-PCCv14 vs RAHT+Pred+Comp BR': -24.60838325895326}]
% [{'ply': 'Data/8iVSLF/Static/loot_viewdep_vox12.ply', 'G-PCCv23 vs RAHT+Pred+Comp BR': -15.676687316890735}]

\begin{figure}[t]
    \centering
    \includegraphics[scale=0.55]{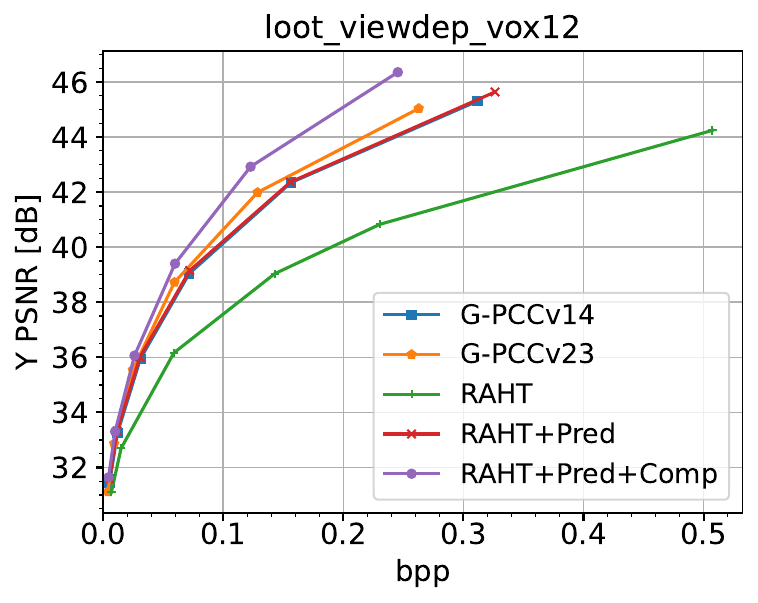}
    \caption{
        Ablation study on loot\_viewdep.
    }
    \label{ab_pred}
\end{figure}
\subsubsection{Ablation Study}
The results of the ablation experiments on loot\_viewdep are presented in Fig.~\ref{ab_pred}. The \texttt{RAHT} represents the vanilla RAHT proposed in \cite{RAHT} without any prediction, which is also the version used by 3DAC. \texttt{RAHT+Pred} refers to DeepRAHT only utilizing the IDW prediction and achieves approximately 48.2\% BD rate savings compared to the vanilla \texttt{RAHT}. \texttt{RAHT+Pred} has the same structure and algorithm as \texttt{G-PCCv14}; therefore, its performance is very close to that of \texttt{G-PCCv14}, which serves as the lower bound of DeepRAHT, ensuring its robustness. DeepRAHT with the prediction compensation module, denoted as \texttt{RAHT+Pred+Comp}, further obtained a 24.6\% BD rate gain compared to \texttt{G-PCCv14}. Additionally, DeepRAHT outperforms \texttt{G-PCCv23} by 16.6\% without requiring sibling context in prediction, as discussed in the \textbf{Prediction Model} section. These results demonstrate the effectiveness of the prediction model and the prediction compensation module of DeepRAHT.

\section{Conclusion}
In this paper, we study the learning of prediction in RAHT. We implemented an end-to-end differentiable predictive RAHT called DeepRAHT, enabling the entire framework to be trained jointly. We propose a learning-based in-loop prediction model that leverages context from the father and grandfather scales, which obtained significant performance gains. We also design a rate proxy based on the Laplace distribution, which has been proven useful for estimating the bitrate of run-length coding and is more efficient and robust for coding RAHT coefficients. DeepRAHT achieves a 16\% bitrate saving compared to G-PCCv23 and outperforms existing learning-based compression methods across multiple datasets, showcasing significant advantages in high performance, low complexity, and exceptional robustness. Future work will focus on extending this approach to the compression of LiDAR and dynamic point clouds.
\section{Acknowledgements} This work is partially supported by the Research Grant Council (RGC) of Hong Kong General Research Fund (GRF) under Grant 11200323, the NSFC/RGC JRS Project N\_CityU198/24, and ITC grant GHP/044/21SZ.

\bibliography{aaai2026}
\end{document}

\maketitle
\setcounter{equation}{24}
\setcounter{figure}{8}
\appendix
\section{Proof of DC Equivalent}
We utilize Enq.~\eqref{Eq4} to efficiently generate the DC of the next scale through sum-pooling. Below is the proof. Following the definition in \textbf{Transform Model}, we only focus on the low-frequency transformation coefficients, and suppose that nodes $g_1,g_2$ are merged into $g_{\text{L}_1}$, and nodes $g_3,g_4$ are merged into $g_{\text{L}_2}$, and so on. From Eqn.~\eqref{eq3}, 
\begin{equation}
\begin{aligned}
g_{\text{L}_1} &= \frac{\sqrt{w_{1}}}{\sqrt{w_{1}+w_{2}}}g_{1} + \frac{\sqrt{w_{2}}}{\sqrt{w_{1}+w_{2}}}g_{2}, \\
g_{\text{L}_2} &= \frac{\sqrt{w_{3}}}{\sqrt{w_{3}+w_{4}}}g_{3} + \frac{\sqrt{w_{4}}}{\sqrt{w_{3}+w_{4}}}g_{4}. \\
\end{aligned}
\end{equation}
From Eqn.~\eqref{eq13}, the weights in nodes $g_{\text{L}_1}$ and $g_{\text{L}_2}$ are updated to $w_1+w_2$ and $w_3+w_4$ respectively. Suppose they are merged into $g_{\text{LL}_1}$ and following Eqn.~\eqref{eq3},
\begin{equation}
    \begin{aligned}
    g_{\text{LL}_1} &= \frac{g_{\text{L}_1} \sqrt{w_{1}+w_2}}{\sqrt{w_{1}+w_{2}+w_{3}+w_{4}}}+ \frac{g_{\text{L}_2}\sqrt{w_{3}+w_4}}{\sqrt{w_{1}+w_{2}+w_{3}+w_{4}}} \\
    &=\frac{ \sum_{j=1}^{4}g_{j}\sqrt{w_{j}}}{\sqrt{\sum_{j=1}^{4}{w_{j}}}}.
    \end{aligned}
\end{equation}
Similarly, suppose that nodes $g_5,g_6,g_7,g_8$ are merged into $g_{\text{LL}_2}$, then we have, 
\begin{equation}
    g_{\text{LL}_2} =\frac{ \sum_{j=5}^{8}g_{j}\sqrt{w_{j}}}{\sqrt{\sum_{j=5}^{8}{w_{j}}}}. \\
\end{equation}
The final DC is obtained by merging $g_{\text{LL}_1}$ and $g_{\text{LL}_2}$,
\begin{equation}
    \begin{aligned}
        g_{\text{LLL}} &= \frac{\sqrt{\sum_{j=1}^4w_j}}{\sum_{j=1}^8\sqrt{w_j}}g_{\text{LL}_1} + \frac{\sqrt{\sum_{j=5}^8w_j}}{\sum_{j=1}^8\sqrt{w_j}}g_{\text{LL}_2} \\
        &=\frac{\sum_{j=1}^{8}g_{j}\sqrt{w_{j}}}{\sqrt{\sum_{j=1}^{8}{w_{j}}}}.\\
    \end{aligned}
    \label{eq4}
\end{equation}
Note that $g_j\equiv A_j/\sqrt{w_j}$, with layer index Eqn.~\eqref{eq4} yields,
\begin{equation}
    \begin{aligned}
        g_{\text{LLL}_{m,{\mathcal{N}_i}}} & = \frac{\sum_{j=1}^{8}A_{{m},j}}{\sqrt{\sum_{j=1}^{8}{w_{j}}}} = \frac{A_{{m+1},i}}{\sqrt{w_{{m+1},i}}}=g_{m+1,i}, \\
    \end{aligned}
\end{equation}
where $w_{{m+1},i}=\sum_{j=1}^8 w_{{m},j}$, $A_{{m+1},i}=\sum_{j=1}^8 A_{{m},j}$.

\section{Zero Run-Length Coding}
Zero run-length coding~\cite{Run_length} is a variant of run-length coding, which is a lossless compression method that efficiently encodes sequences of data by storing the (\texttt{run}, \texttt{value}) entries, indicating the lengths of continuous zeros (\texttt{run}) and non-zero values (\texttt{value}). For example, if the input sequence is A000BC0D, the corresponding zero run-length code is 0A3B0C1D. The \texttt{value} is binarized by exponential-Golomb code, and the \texttt{run} is binarized by a cascade of unary code, truncated Rice code, and exponential-Golomb code. After the binarization, the codewords are compressed by arithmetic coding.

\begin{figure}[]
    \centering
    \includegraphics[scale=0.6]{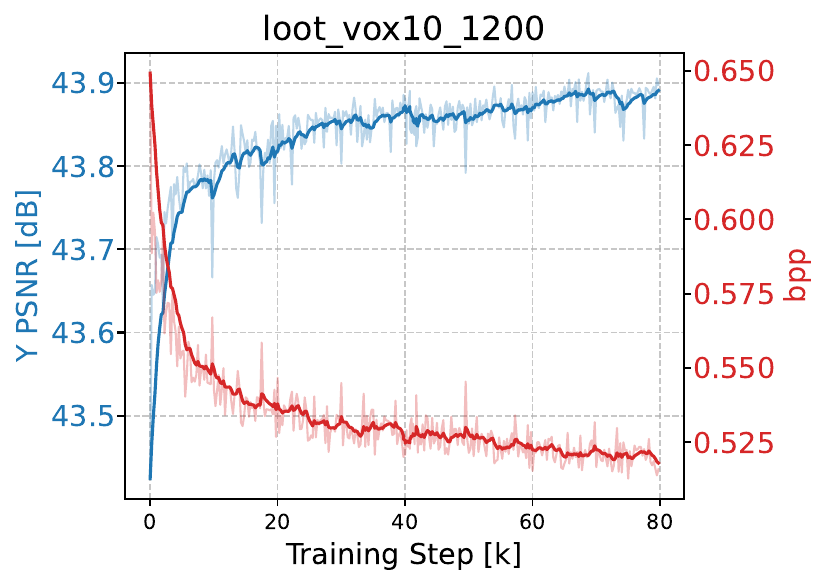}
    \caption{
       Y PSNR and bitrate of validation data.
    }
    \label{training_process}
\end{figure}
\section{Training Process}
The validation loss tested on loot\_vox10 during the training process is shown in Fig.~\ref{training_process}. It can be observed that both the PSNR and bitrate are well initialized at the beginning of the training. This verifies the performance for DeepRAHT without training. Meanwhile, as the training progresses, the reconstruction quality of the validation data improves and the bitrate decreases, which proves the effectiveness of the loss function and the rate proxy. In contrast, 3DAC is only an entropy model of the RAHT coefficients, so the reconstruction quality during its training process does not change.

\onecolumn
\section{R-D Curves Details}
The R-D curves of each testing data are presented below.

\begin{figure*}[htb]
    \centering
    \begin{adjustbox}{max width=\textwidth} 
        \subfigure{
            \includegraphics[width=0.29\textwidth]{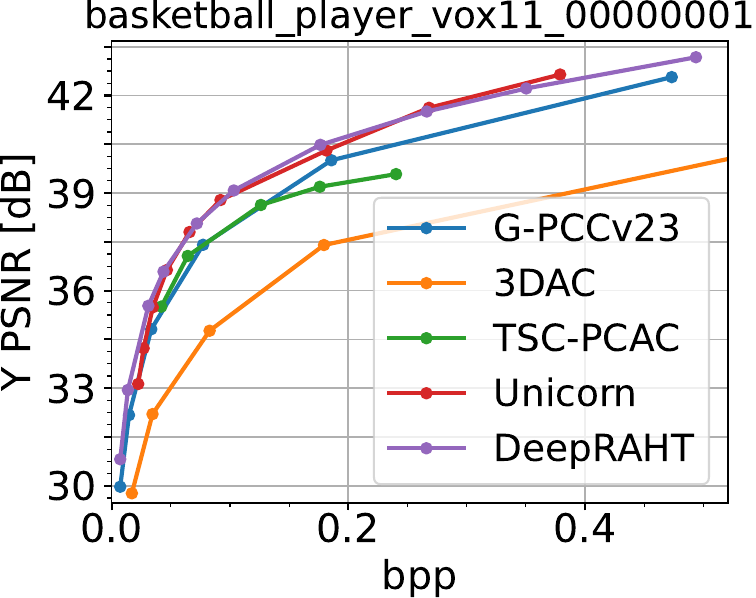}
        }
        \hfill
        \subfigure{
            \includegraphics[width=0.29\textwidth]{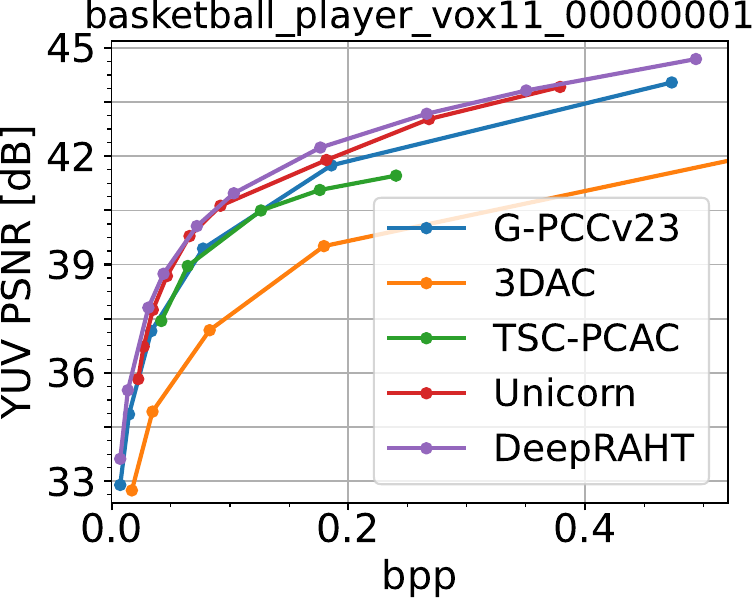}
        }
        \subfigure{
            \includegraphics[width=0.28\textwidth]{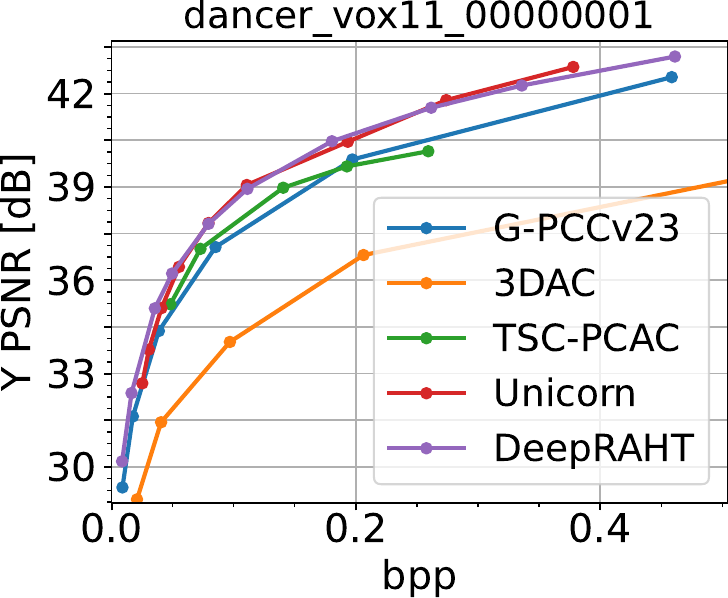}
        }
        \hfill
        \subfigure{
            \includegraphics[width=0.28\textwidth]{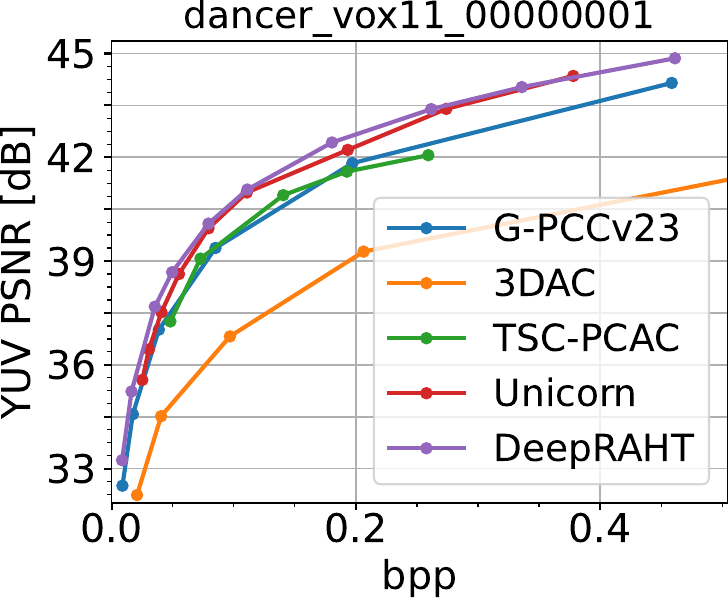}
        }
    \end{adjustbox}
    \begin{adjustbox}{max width=\textwidth} 

        \subfigure{
            \includegraphics[width=0.28\textwidth]{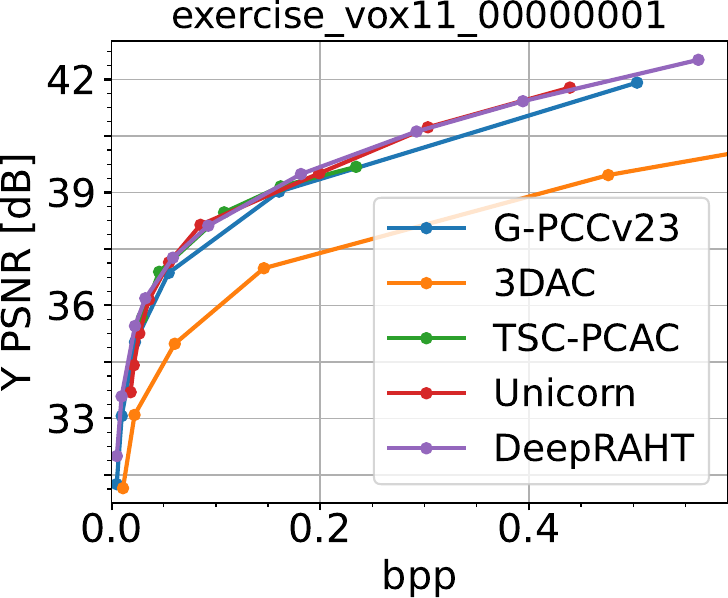}
        }
        \hfill
        \subfigure{
            \includegraphics[width=0.28\textwidth]{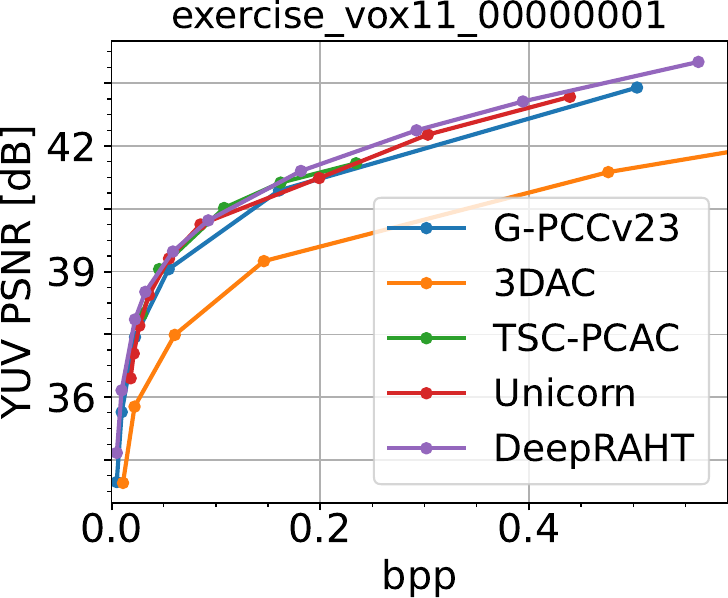}
        }
        \subfigure{
            \includegraphics[width=0.28\textwidth]{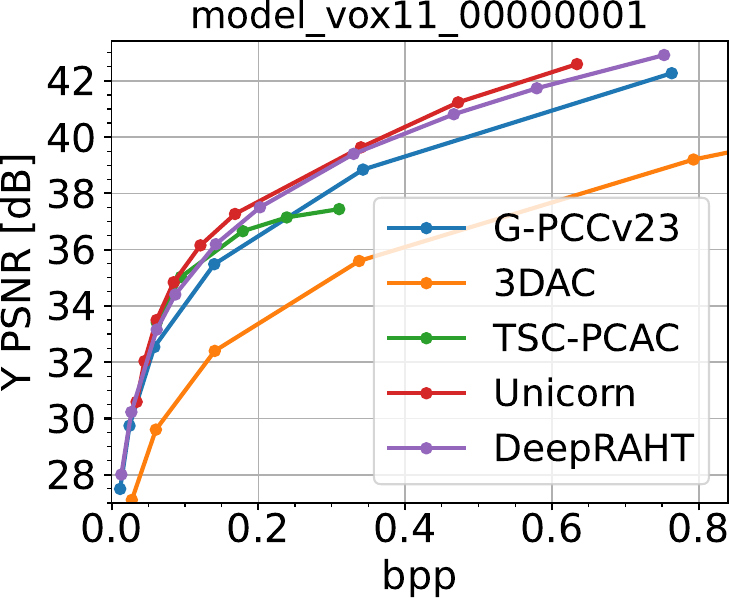}
        }
        \hfill
        \subfigure{
            \includegraphics[width=0.28\textwidth]{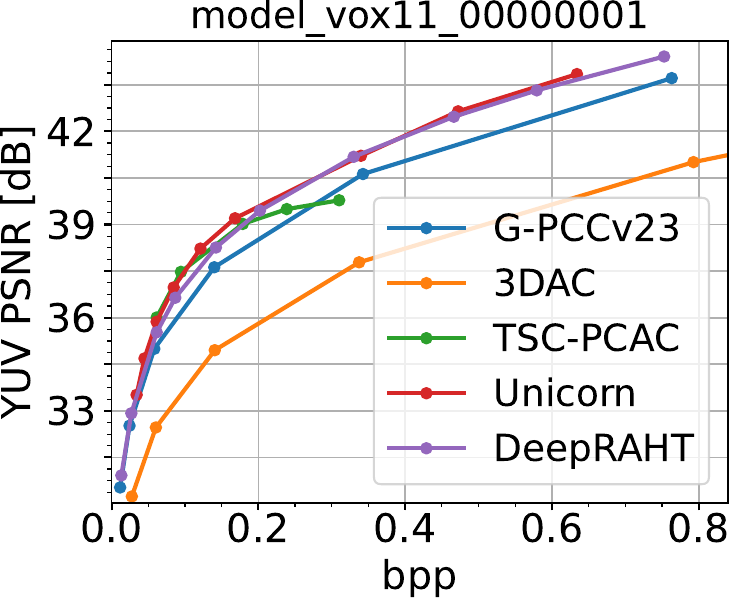}
        }
    \end{adjustbox}
\caption[]{Detailed R-D comparison of Owlii dataset.}
\end{figure*}

\begin{figure*}[!h]
    \centering
    \begin{adjustbox}{max width=\textwidth} 

        \subfigure{
            \includegraphics[width=0.28\textwidth]{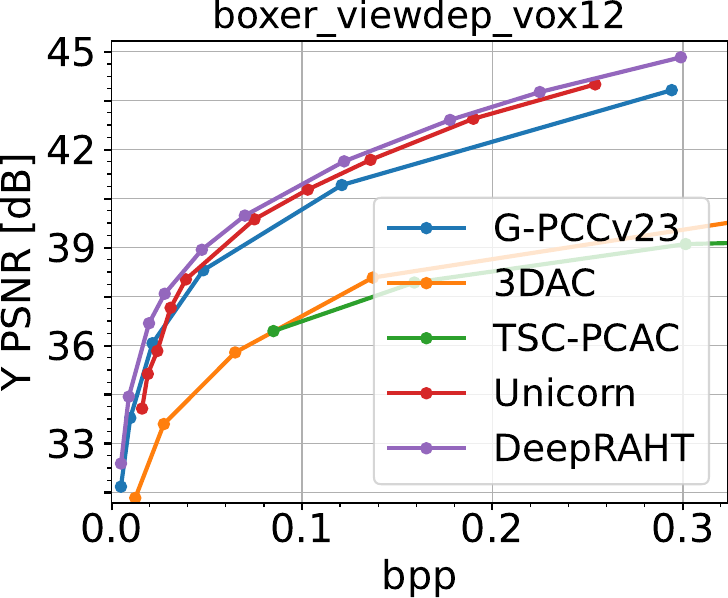}
        }
        \hfill
        \subfigure{
            \includegraphics[width=0.28\textwidth]{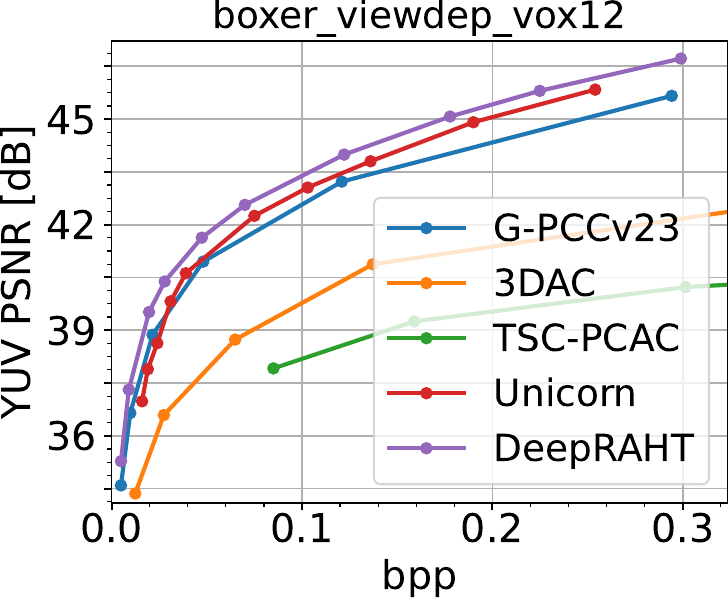}
        }
        \subfigure{
            \includegraphics[width=0.28\textwidth]{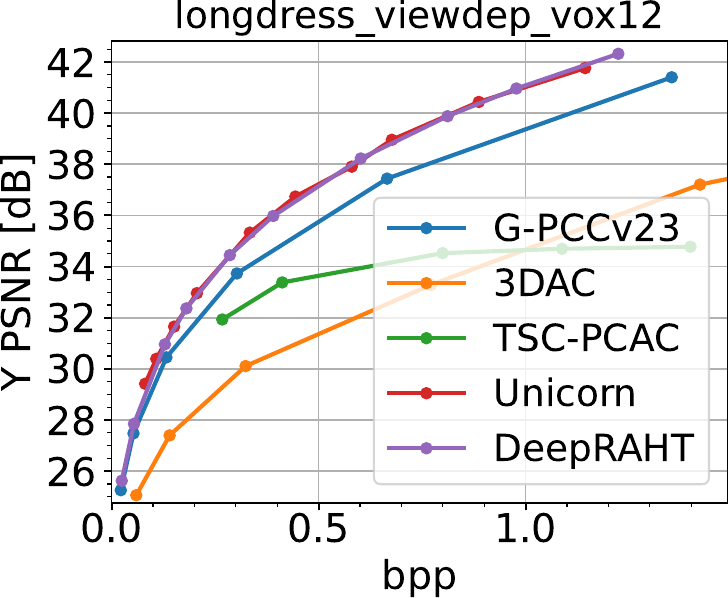}
        }
        \hfill
        \subfigure{
            \includegraphics[width=0.28\textwidth]{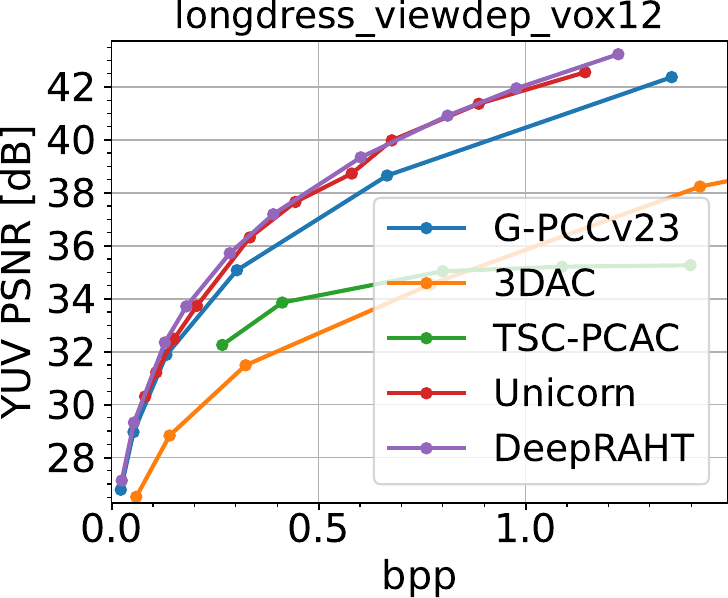}
        }
    \end{adjustbox}
    \begin{adjustbox}{max width=\textwidth} 

        \subfigure{
            \includegraphics[width=0.28\textwidth]{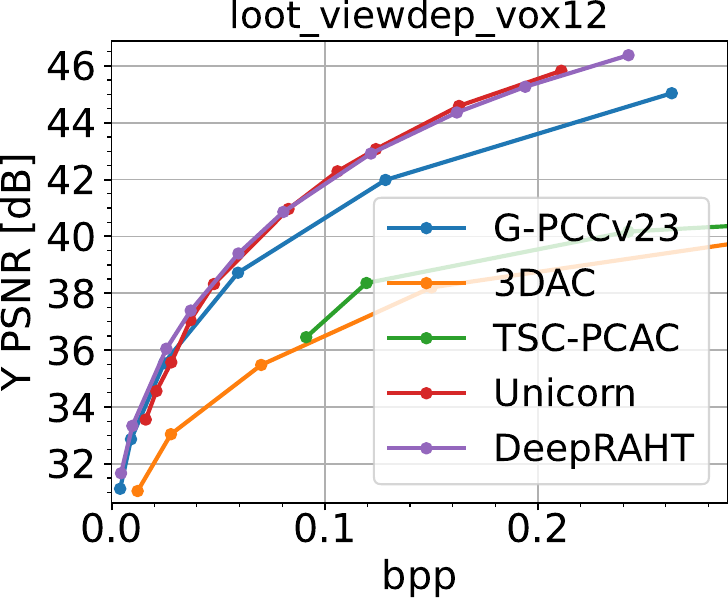}
        }
        \hfill
        \subfigure{
            \includegraphics[width=0.28\textwidth]{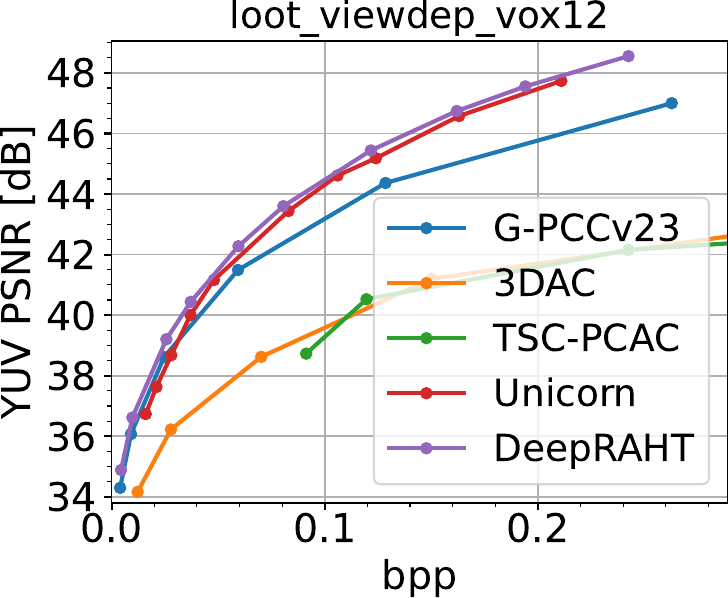}
        }

        \subfigure{
            \includegraphics[width=0.28\textwidth]{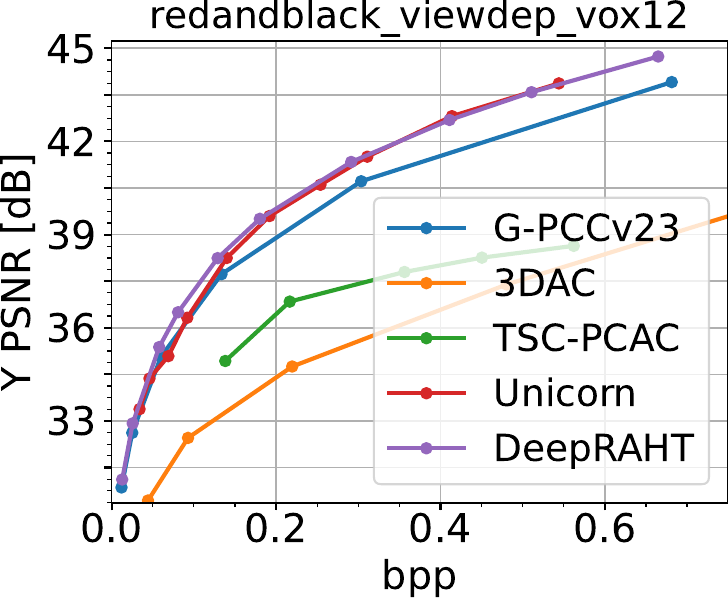}
        }
        \hfill
        \subfigure{
            \includegraphics[width=0.28\textwidth]{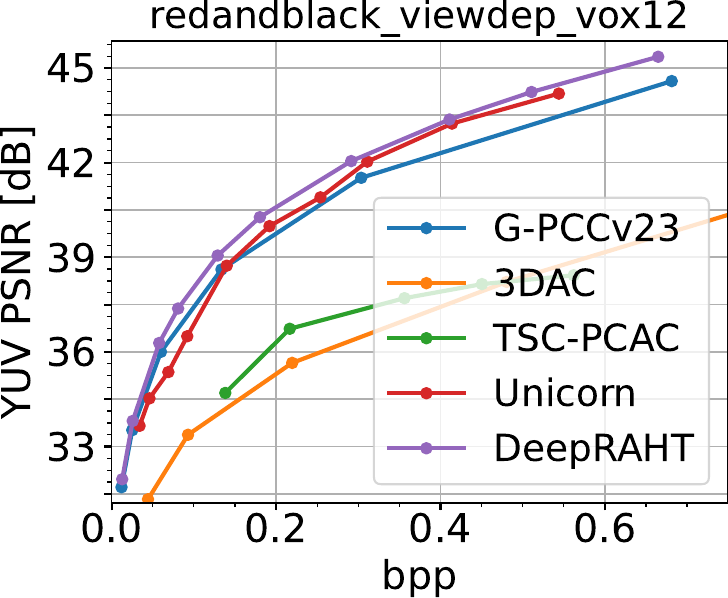}
        }
    \end{adjustbox}
    \begin{adjustbox}{max width=\textwidth} 

        \subfigure{
            \includegraphics[width=0.28\textwidth]{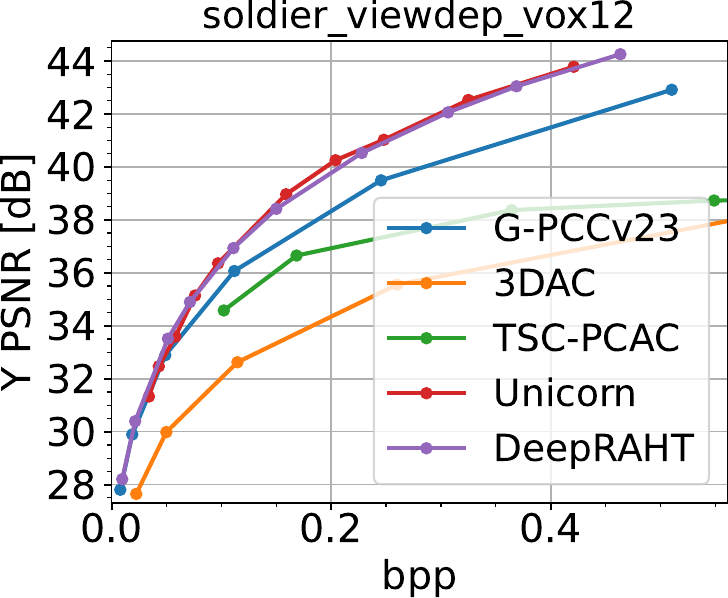}
        }
        \hfill
        \subfigure{
            \includegraphics[width=0.28\textwidth]{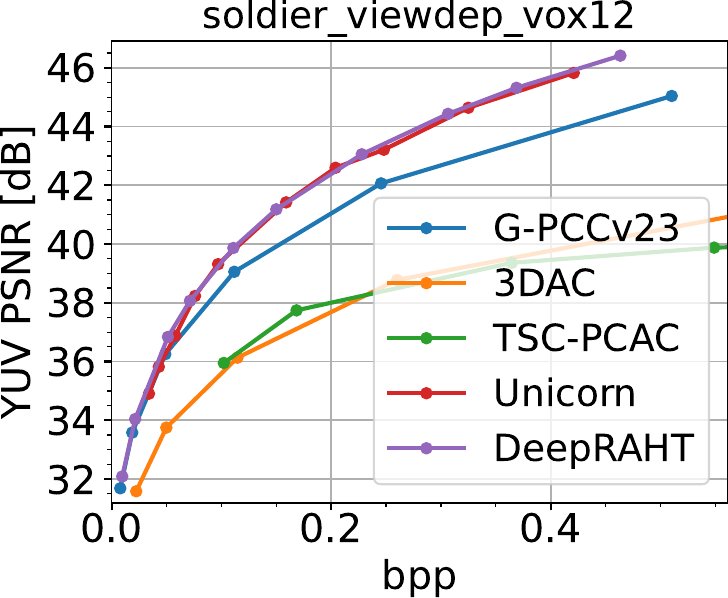}
        }

        \subfigure{
            \includegraphics[width=0.28\textwidth]{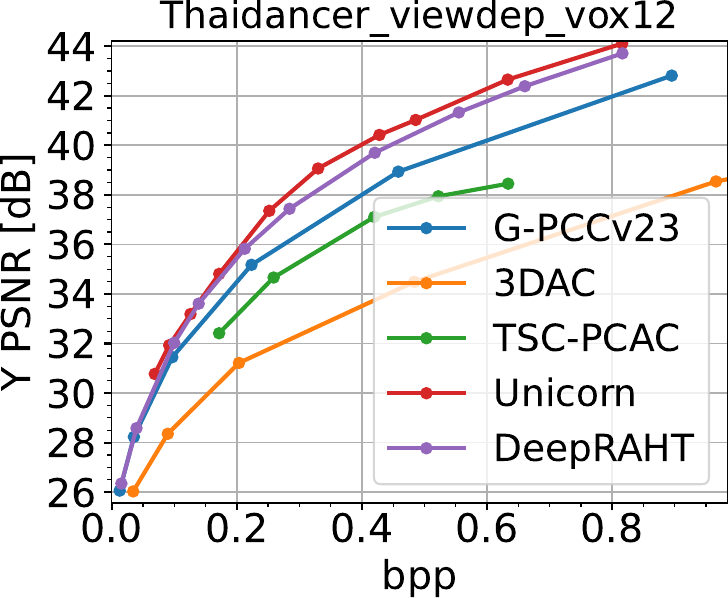}
        }
        \hfill
        \subfigure{
            \includegraphics[width=0.28\textwidth]{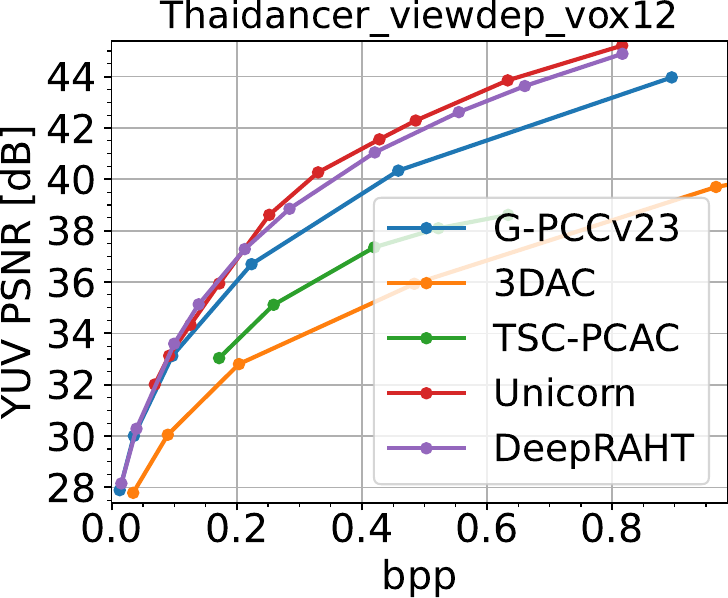}
        }
    \end{adjustbox}
\caption[]{Detailed R-D comparison of 8iVSLF dataset.}
\end{figure*}

\begin{figure*}[!t]
    \centering
    \vspace{-5cm}
        \begin{adjustbox}{max width=\textwidth} 
    
            \subfigure{
                \includegraphics[width=0.28\textwidth]{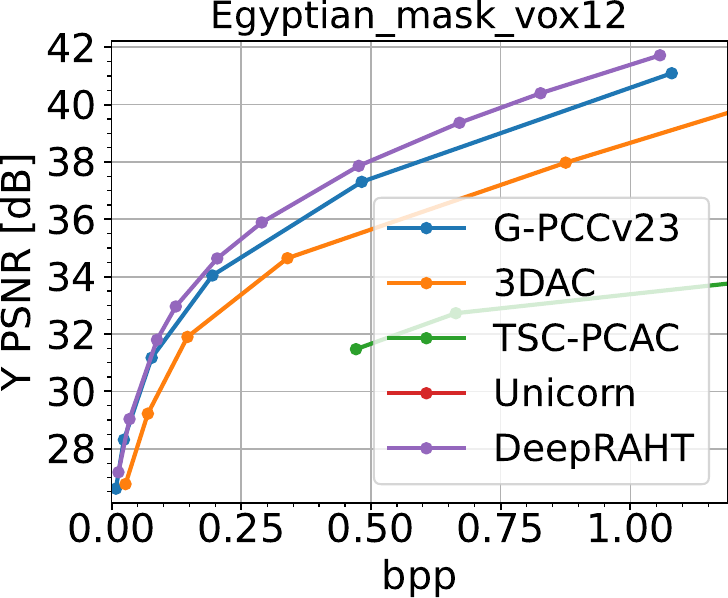}
            }
            \hfill
            \subfigure{
                \includegraphics[width=0.28\textwidth]{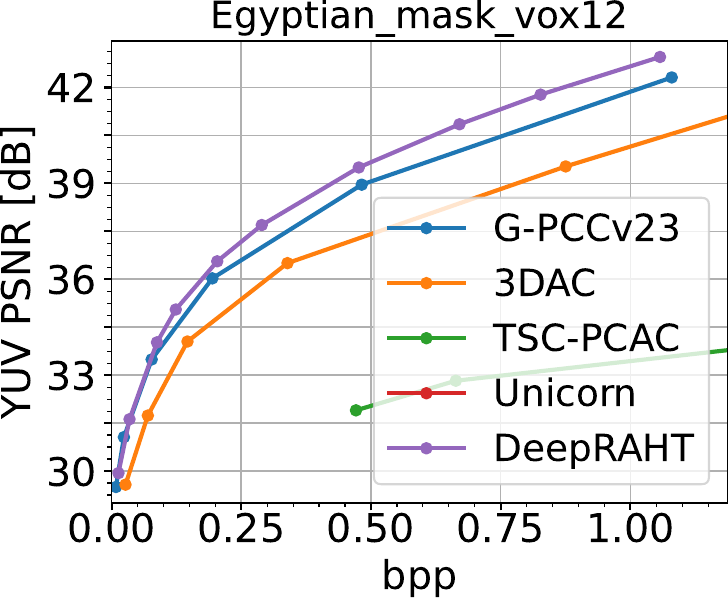}
            }
    
            \subfigure{
                \includegraphics[width=0.28\textwidth]{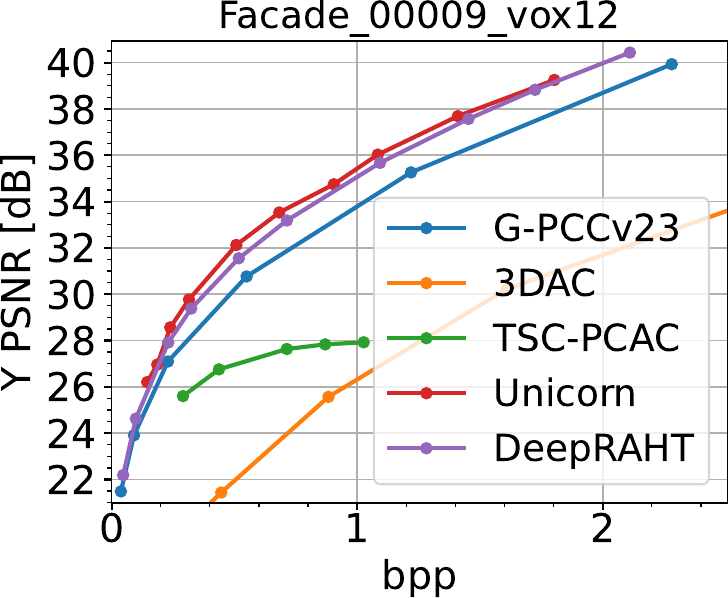}
            }
            \hfill
            \subfigure{
                \includegraphics[width=0.28\textwidth]{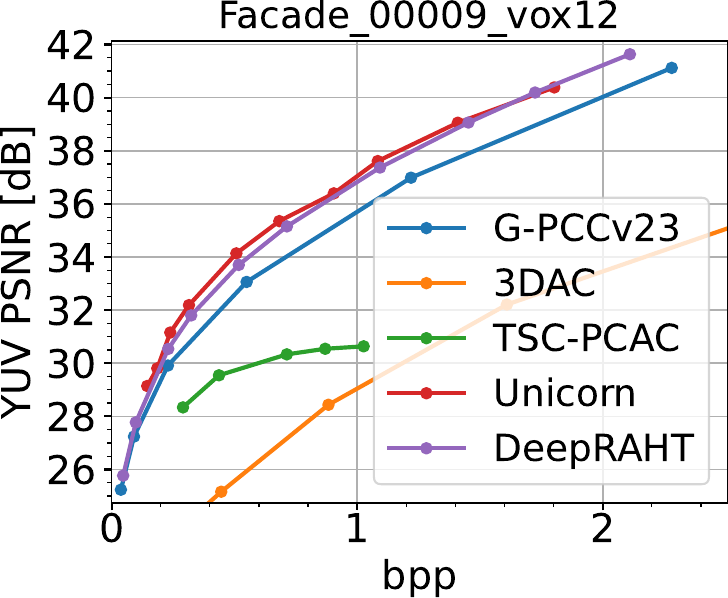}
            }
        \end{adjustbox}
        \begin{adjustbox}{max width=\textwidth} 
    
            \subfigure{
                \includegraphics[width=0.28\textwidth]{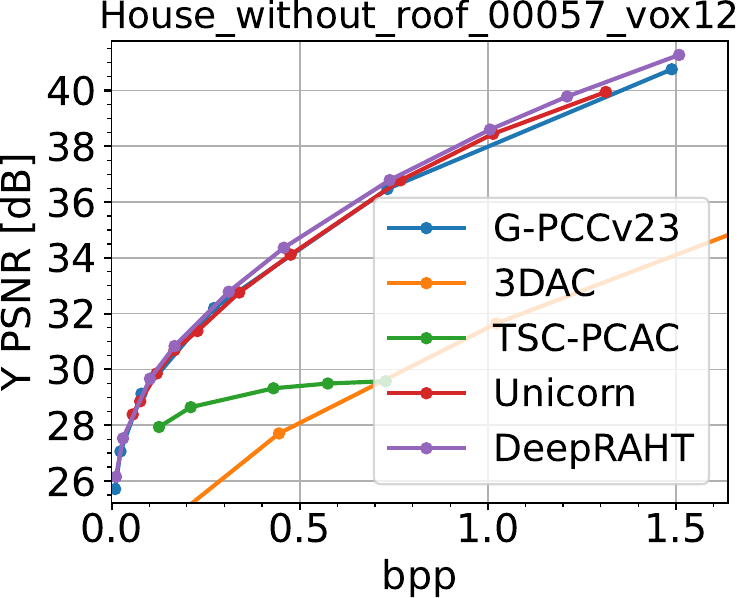}
            }
            \hfill
            \subfigure{
                \includegraphics[width=0.28\textwidth]{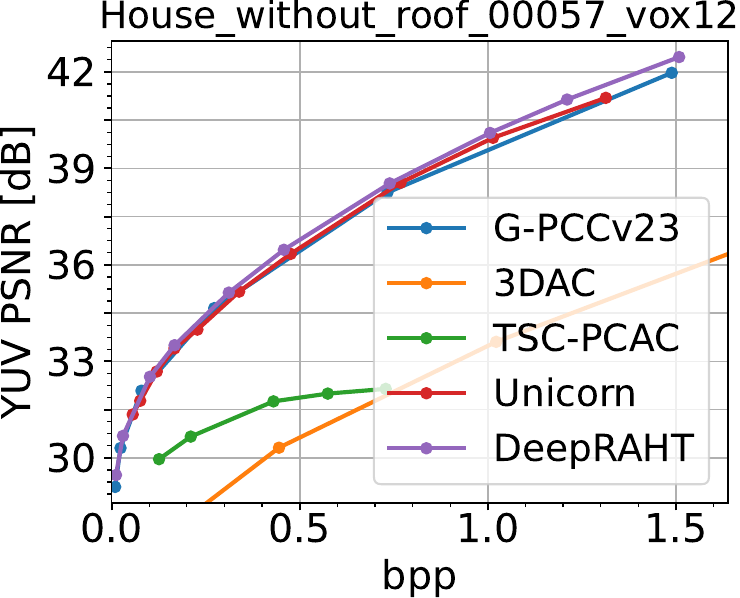}
            }
    
            \subfigure{
                \includegraphics[width=0.28\textwidth]{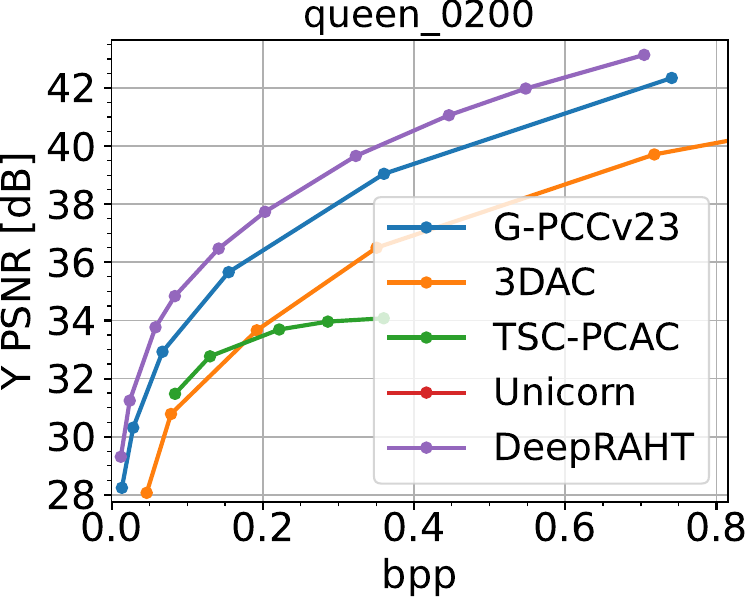}
            }
            \hfill
            \subfigure{
                \includegraphics[width=0.28\textwidth]{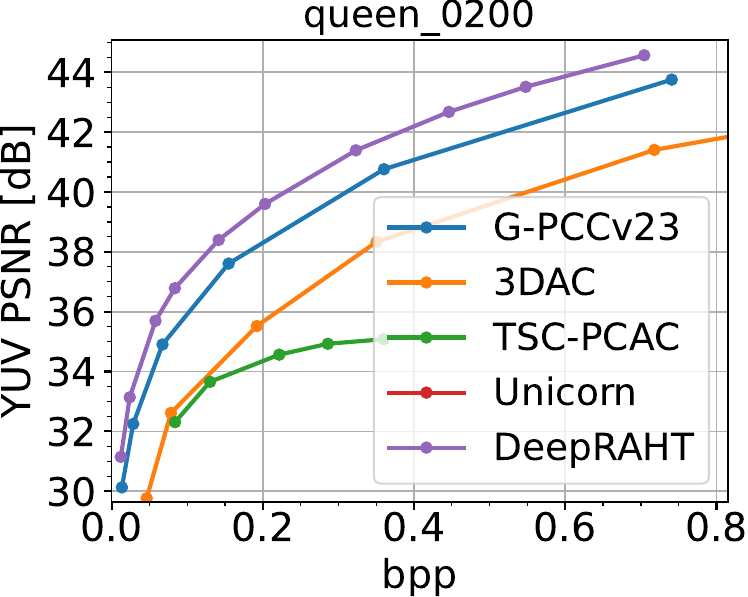}
            }
        \end{adjustbox}
        \begin{adjustbox}{max width=\textwidth} 
    
            \subfigure{
                \includegraphics[width=0.28\textwidth]{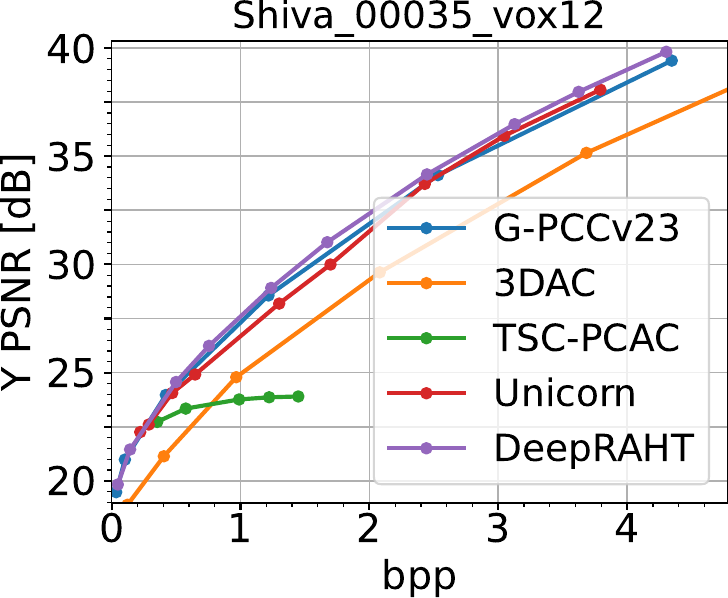}
            }
            \hfill
            \subfigure{
                \includegraphics[width=0.28\textwidth]{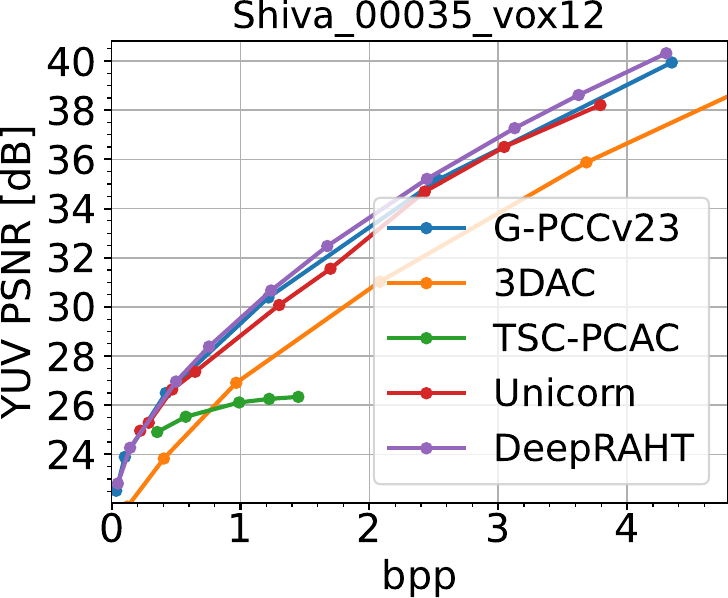}
            }
    
            \subfigure{
                \includegraphics[width=0.28\textwidth]{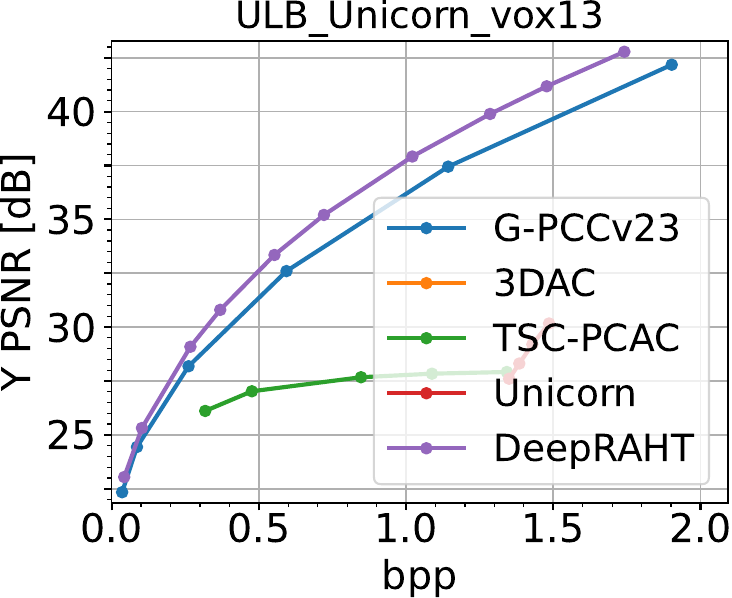}
            }
            \hfill
            \subfigure{
                \includegraphics[width=0.28\textwidth]{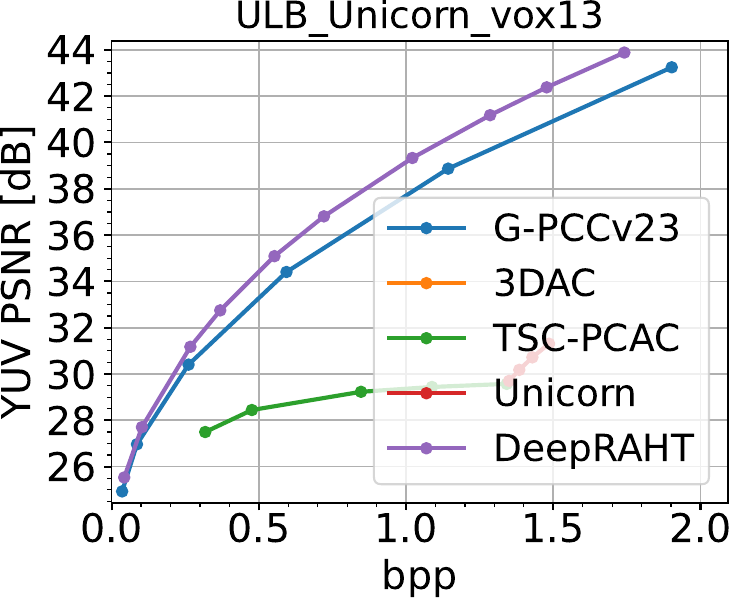}
            }
        \end{adjustbox}
        \begin{adjustbox}{max width=\textwidth} 

        \subfigure{
            \includegraphics[width=0.28\textwidth]{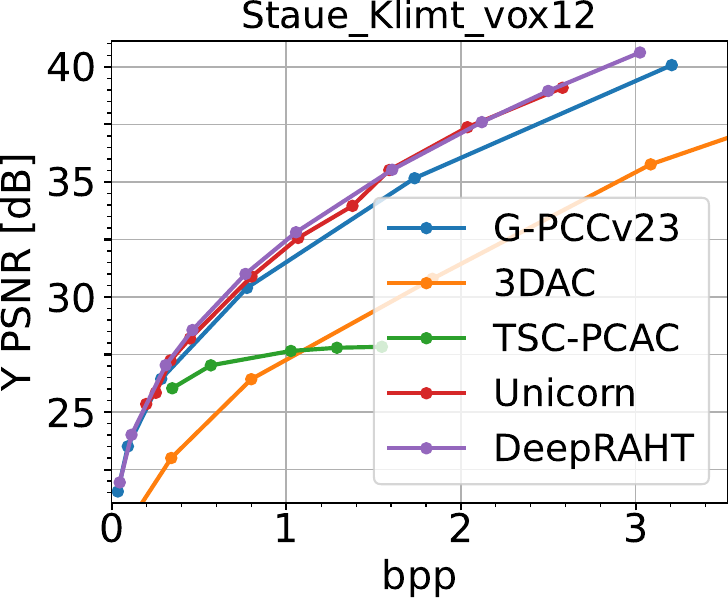}
        }
        \hfill
        \subfigure{
            \includegraphics[width=0.28\textwidth]{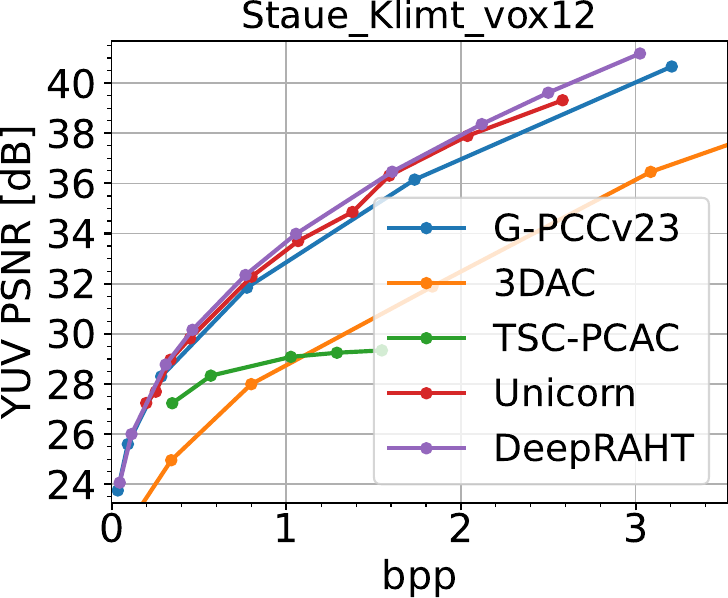}
        }

        \subfigure{
            \textcolor{white}{\rule{0.28\textwidth}{0.2\textwidth}} 
        }
        \hfill
        \subfigure{
            \textcolor{white}{\rule{0.28\textwidth}{0.2\textwidth}} 
        }
        \end{adjustbox}
        \caption[]{Detailed R-D comparison of MPEG samples.}
    \end{figure*}
    
    \twocolumn

    \onecolumn
    \section{Chroma R-D Curves Details}
    The chroma R-D curves of each testing data are presented below.
    
    \begin{figure*}[htb]
        \centering
        \begin{adjustbox}{max width=\textwidth} 
    
            \subfigure{
                \includegraphics[width=0.29\textwidth]{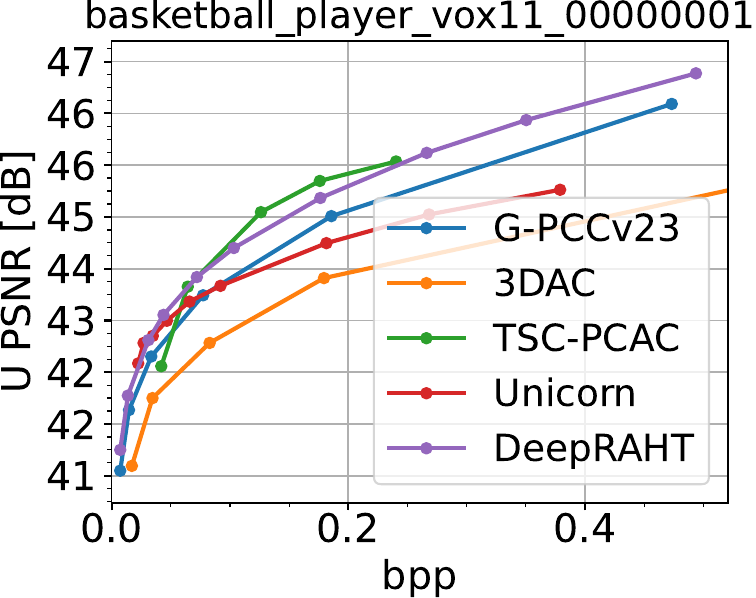}
            }
            \hfill
            \subfigure{
                \includegraphics[width=0.29\textwidth]{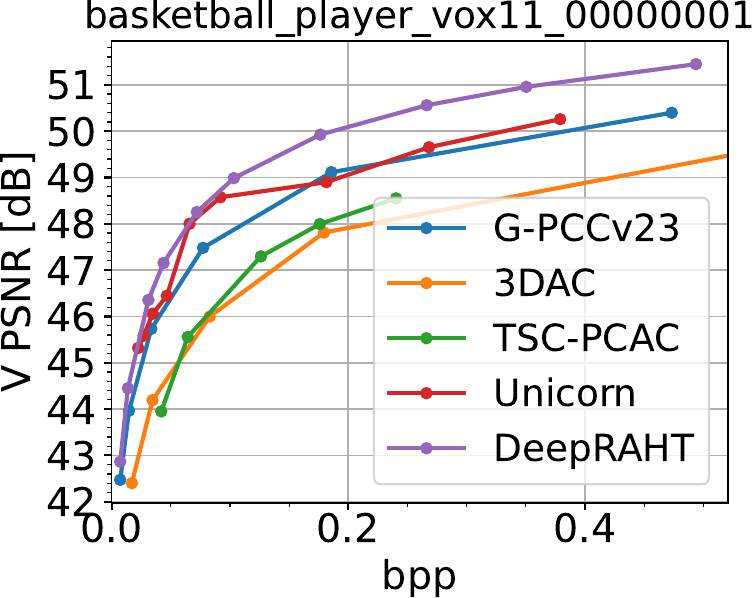}
            }
            \subfigure{
                \includegraphics[width=0.28\textwidth]{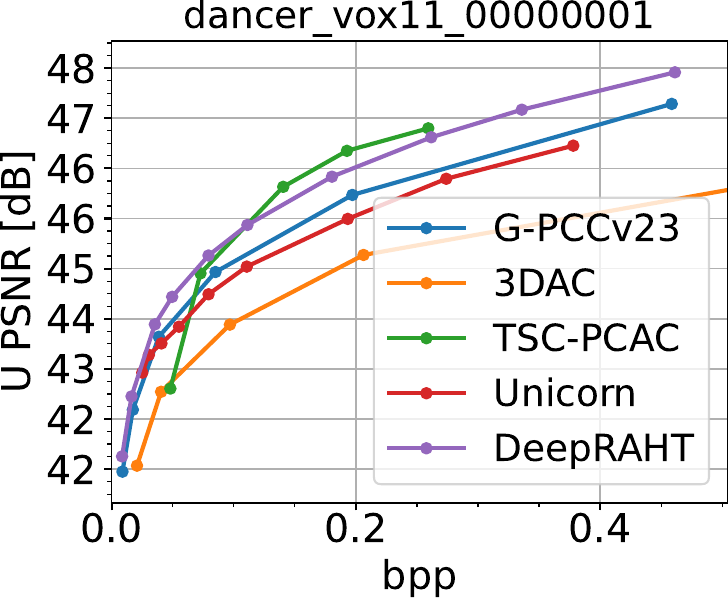}
            }
            \hfill
            \subfigure{
                \includegraphics[width=0.28\textwidth]{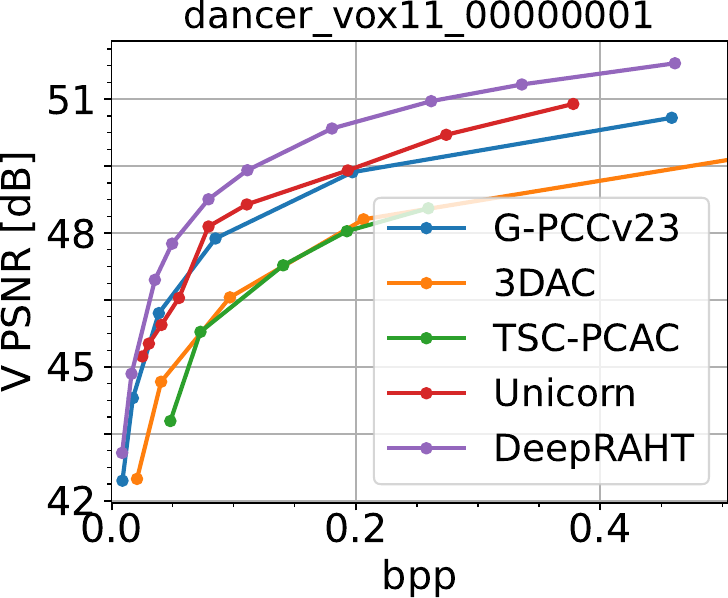}
            }
        \end{adjustbox}
        \begin{adjustbox}{max width=\textwidth} 
    
            \subfigure{
                \includegraphics[width=0.28\textwidth]{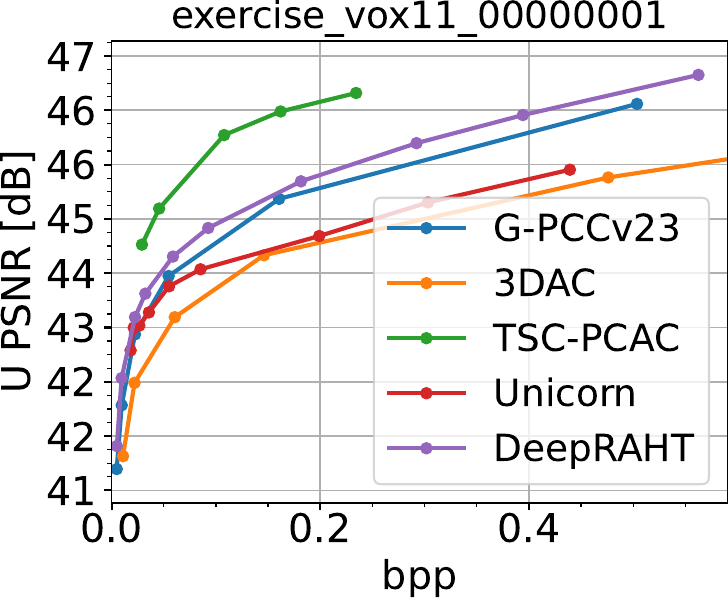}
            }
            \hfill
            \subfigure{
                \includegraphics[width=0.28\textwidth]{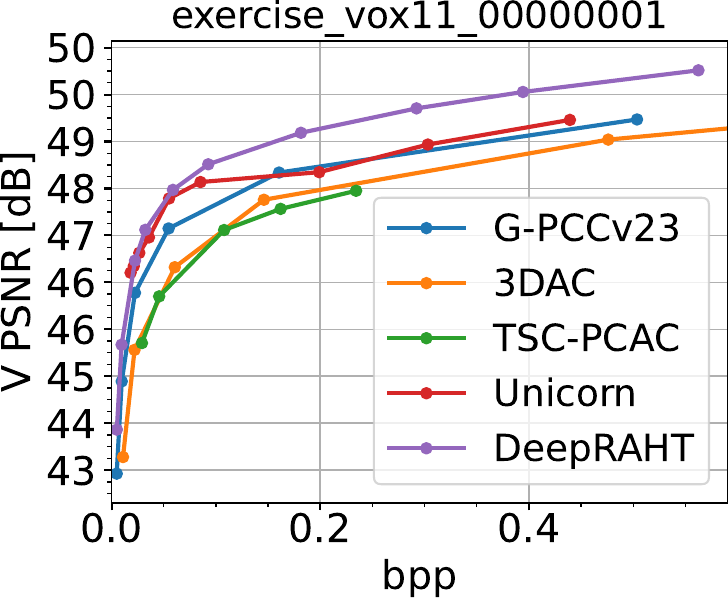}
            }
            \subfigure{
                \includegraphics[width=0.28\textwidth]{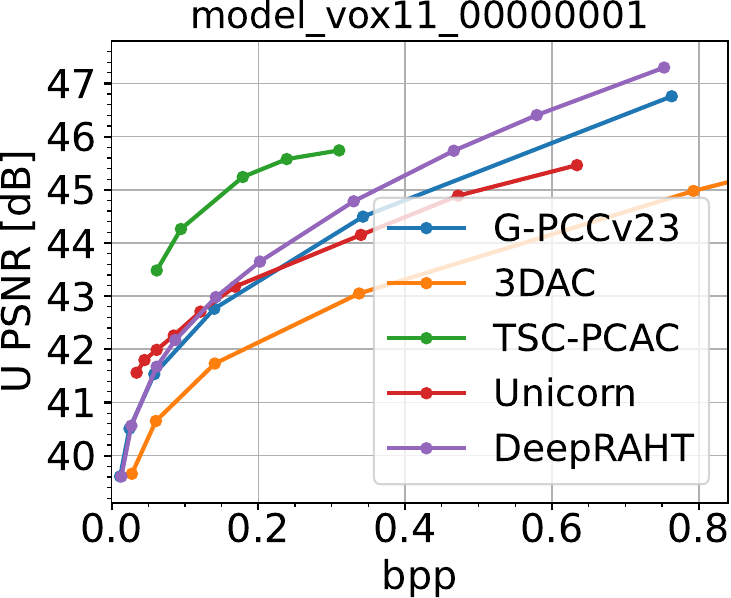}
            }
            \hfill
            \subfigure{
                \includegraphics[width=0.28\textwidth]{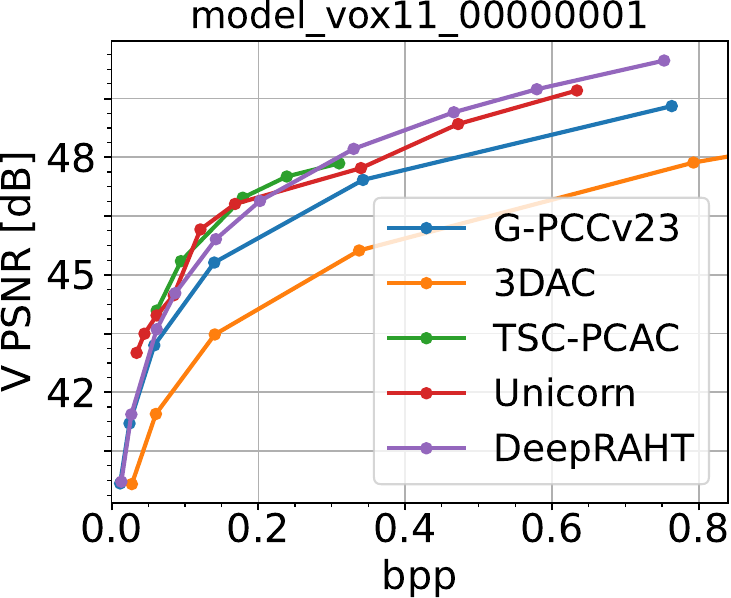}
            }
        \end{adjustbox}
    \caption[]{Detailed chroma R-D comparison of Owlii dataset.}
    \end{figure*}
    
    \begin{figure*}[!h]
        \centering
        \begin{adjustbox}{max width=\textwidth} 
    
            \subfigure{
                \includegraphics[width=0.28\textwidth]{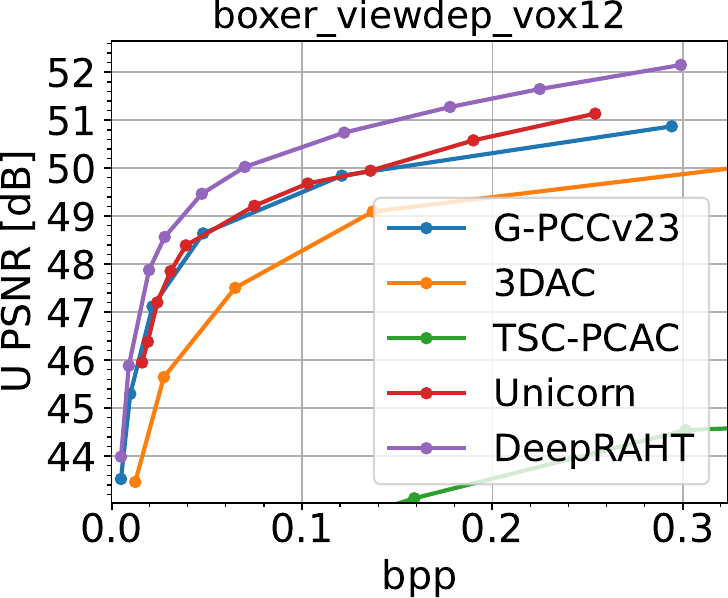}
            }
            \hfill
            \subfigure{
                \includegraphics[width=0.28\textwidth]{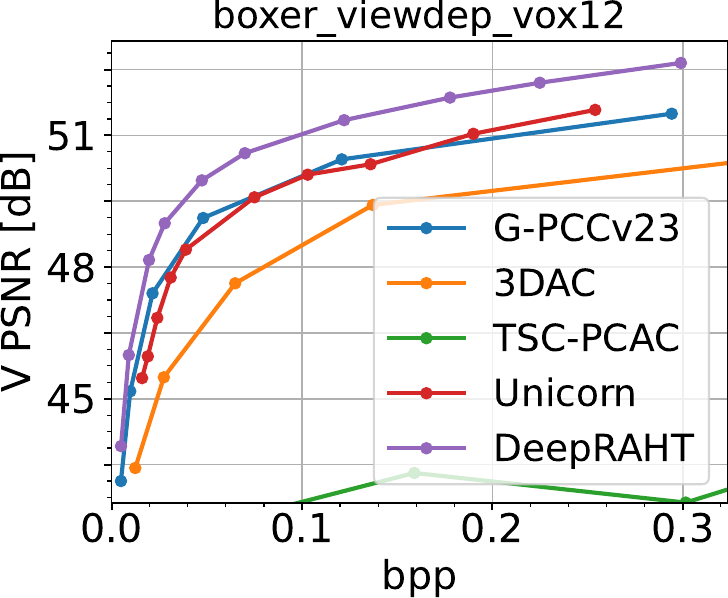}
            }
            \subfigure{
                \includegraphics[width=0.28\textwidth]{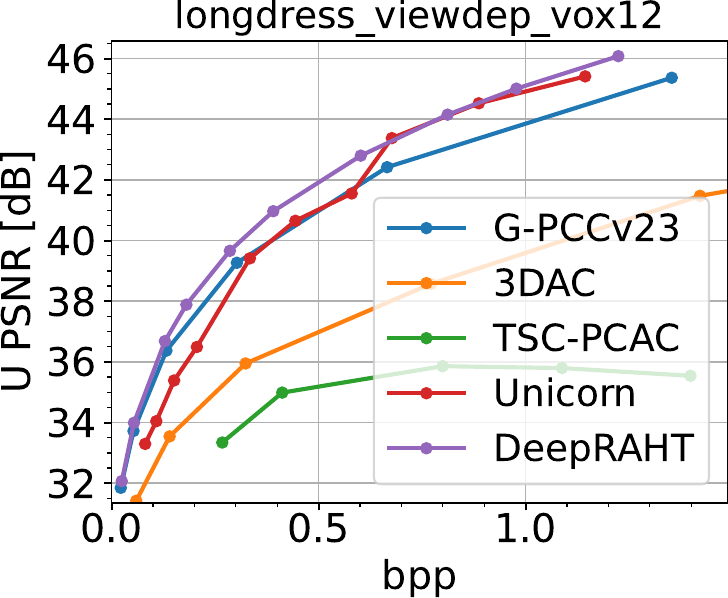}
            }
            \hfill
            \subfigure{
                \includegraphics[width=0.28\textwidth]{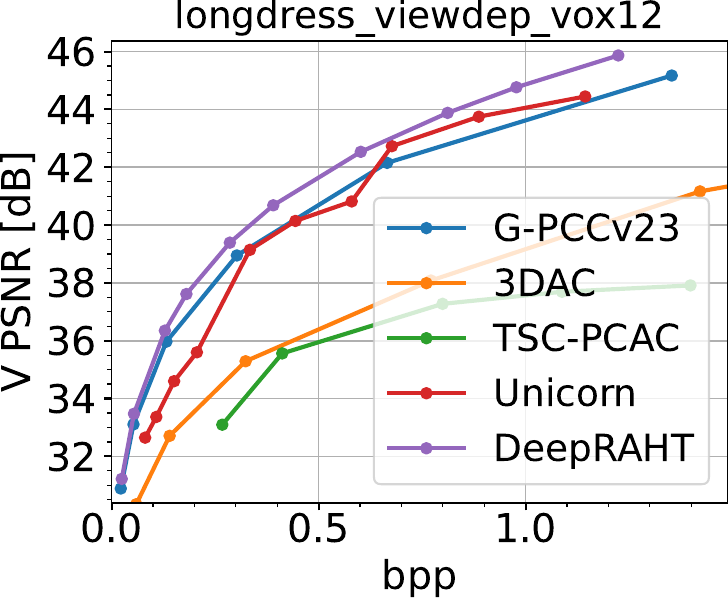}
            }
        \end{adjustbox}
        \begin{adjustbox}{max width=\textwidth} 
    
            \subfigure{
                \includegraphics[width=0.28\textwidth]{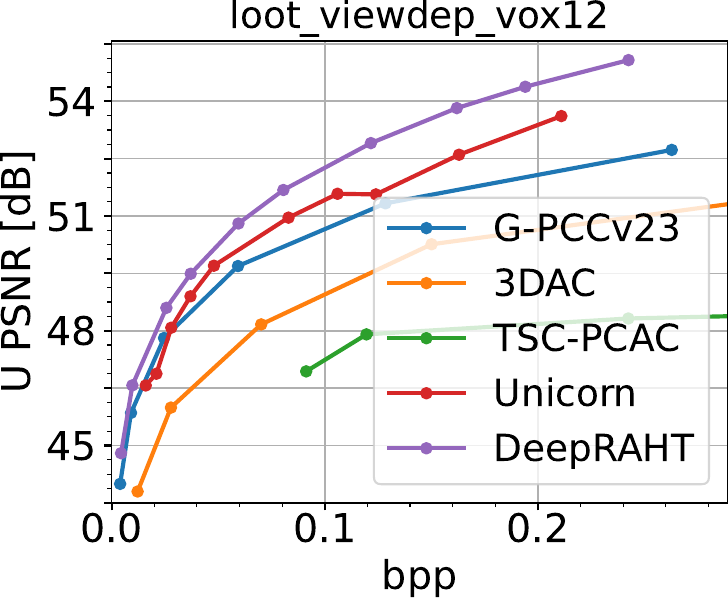}
            }
            \hfill
            \subfigure{
                \includegraphics[width=0.28\textwidth]{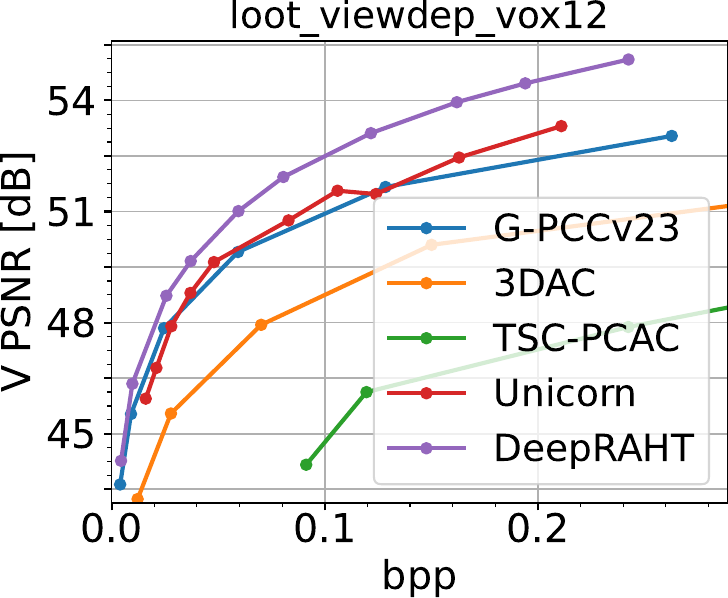}
            }
    
            \subfigure{
                \includegraphics[width=0.28\textwidth]{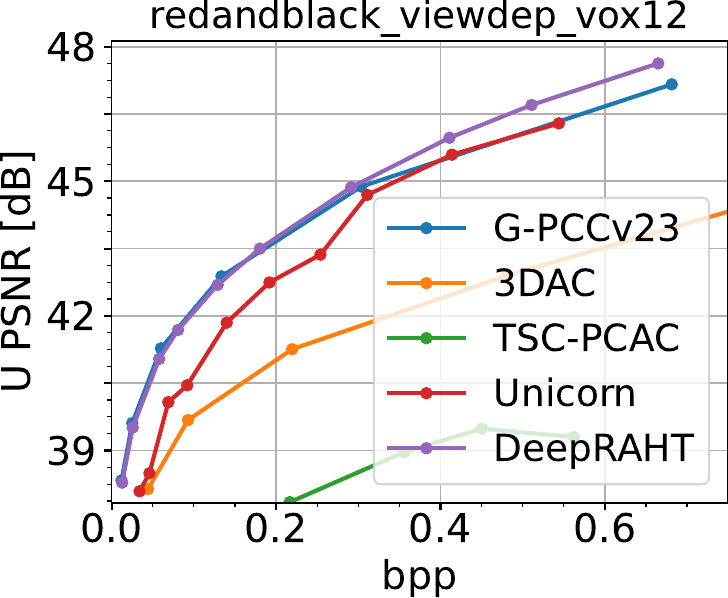}
            }
            \hfill
            \subfigure{
                \includegraphics[width=0.28\textwidth]{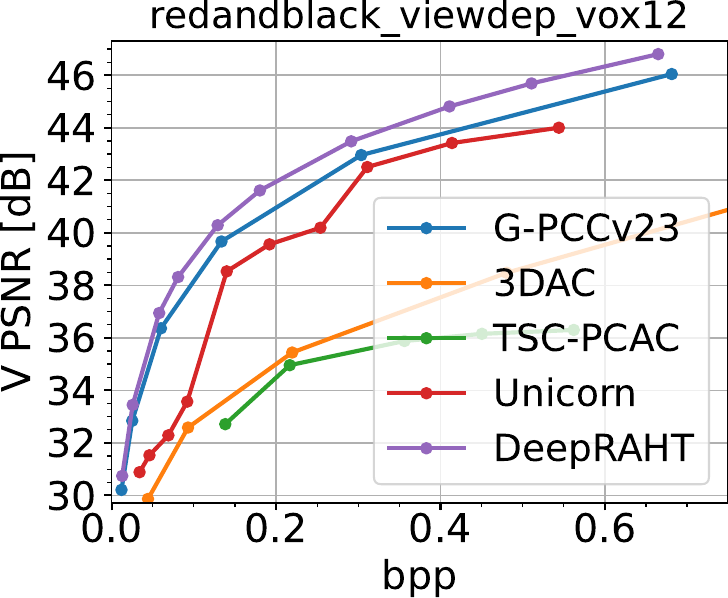}
            }
        \end{adjustbox}
        \begin{adjustbox}{max width=\textwidth} 
    
            \subfigure{
                \includegraphics[width=0.28\textwidth]{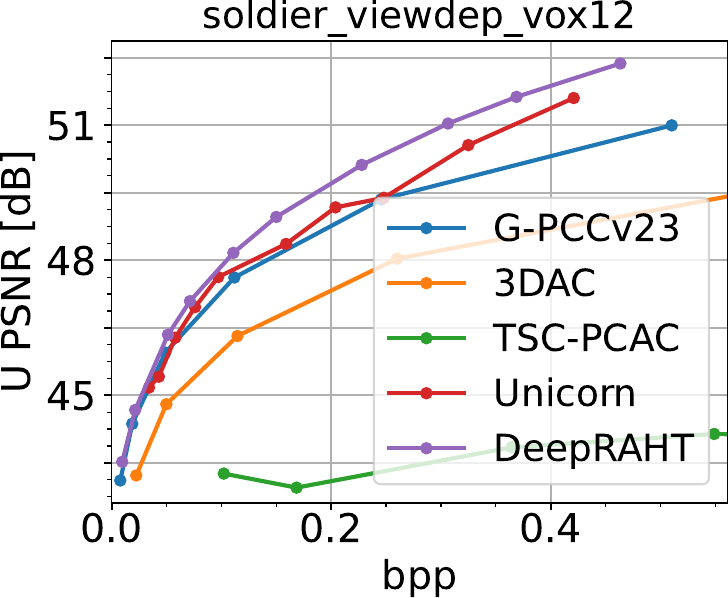}
            }
            \hfill
            \subfigure{
                \includegraphics[width=0.28\textwidth]{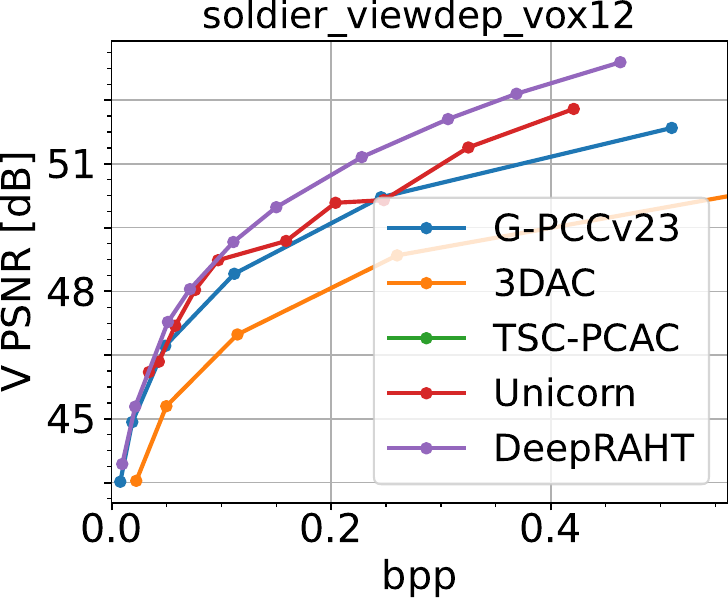}
            }
    
            \subfigure{
                \includegraphics[width=0.28\textwidth]{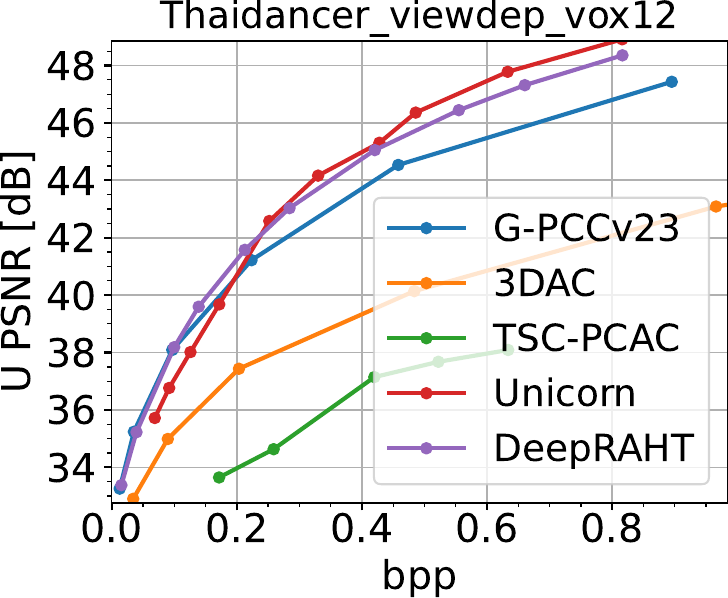}
            }
            \hfill
            \subfigure{
                \includegraphics[width=0.28\textwidth]{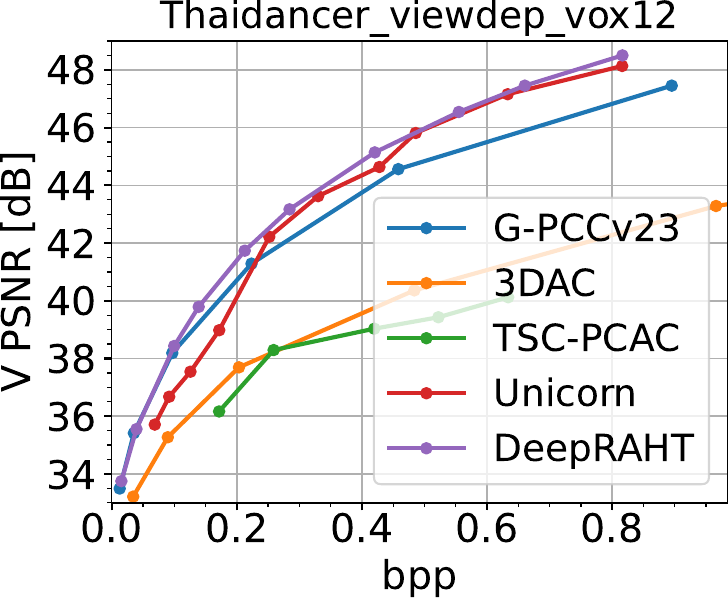}
            }
        \end{adjustbox}
    \caption[]{Detailed chroma R-D comparison of 8iVSLF dataset.}
    \end{figure*}

    \begin{figure*}[!t]
        \centering
        % \vspace{-10cm}
            \begin{adjustbox}{max width=\textwidth} 
        
                \subfigure{
                    \includegraphics[width=0.28\textwidth]{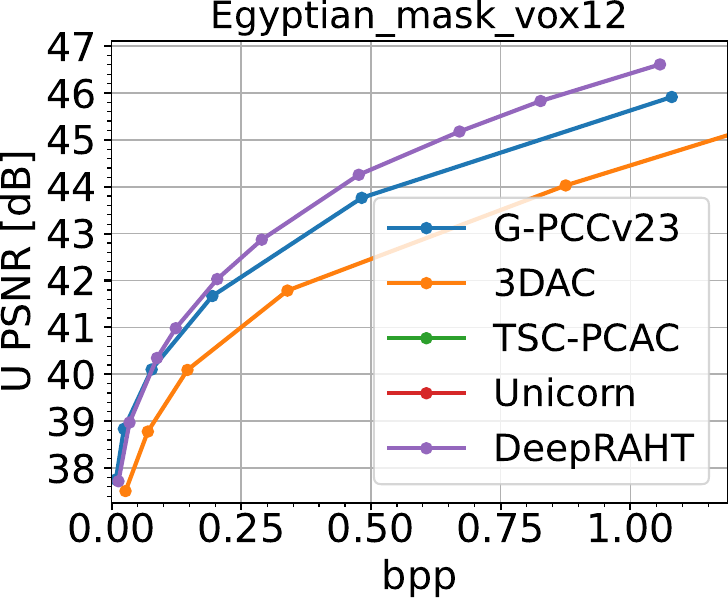}
                }
                \hfill
                \subfigure{
                    \includegraphics[width=0.28\textwidth]{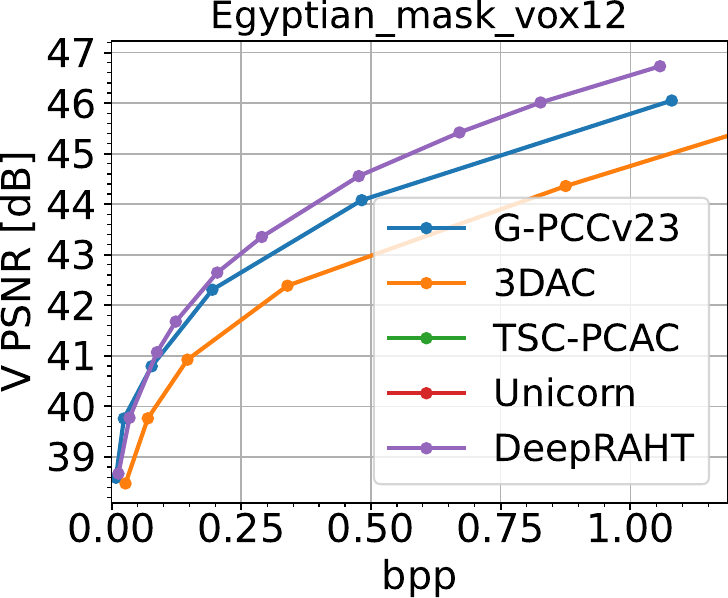}
                }
        
                \subfigure{
                    \includegraphics[width=0.28\textwidth]{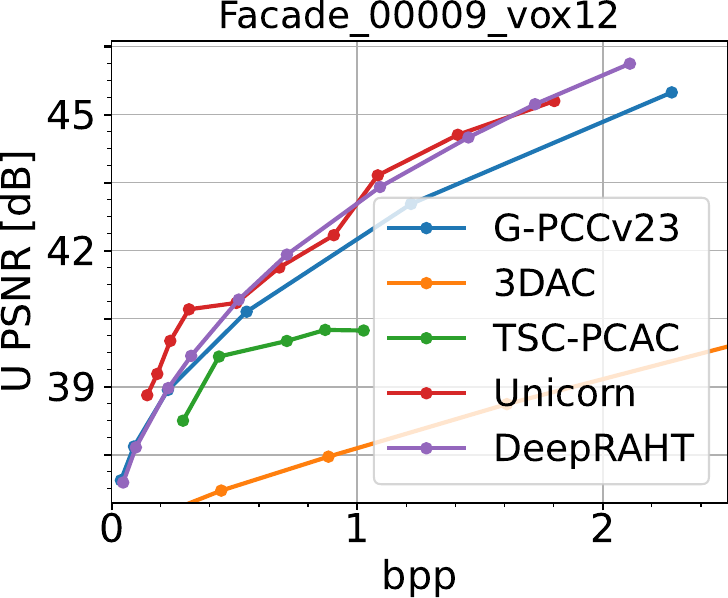}
                }
                \hfill
                \subfigure{
                    \includegraphics[width=0.28\textwidth]{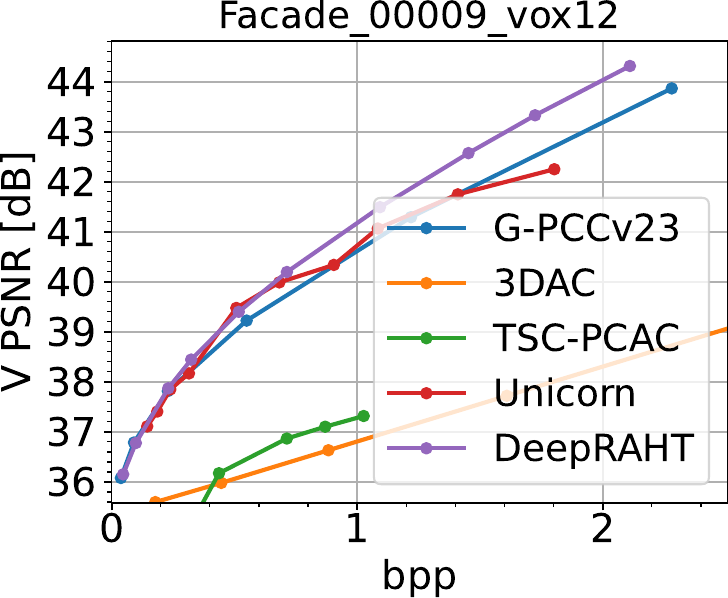}
                }
            \end{adjustbox}
            \begin{adjustbox}{max width=\textwidth} 
        
                \subfigure{
                    \includegraphics[width=0.28\textwidth]{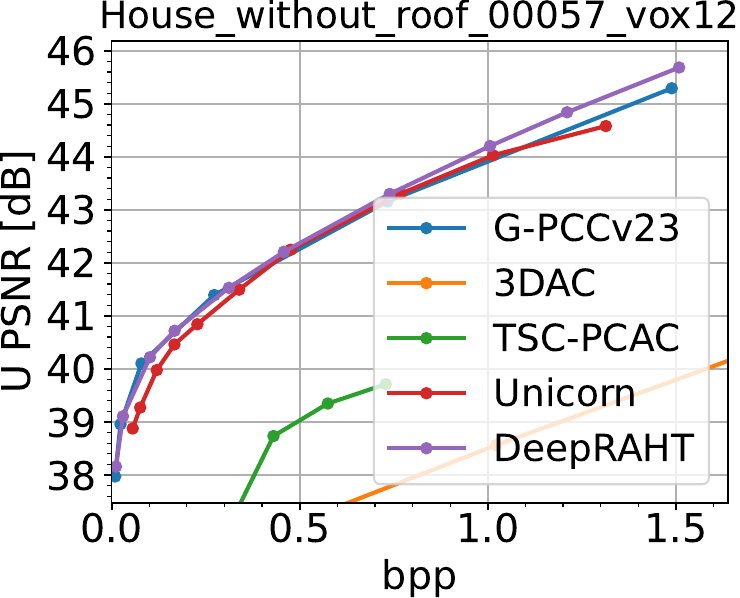}
                }
                \hfill
                \subfigure{
                    \includegraphics[width=0.28\textwidth]{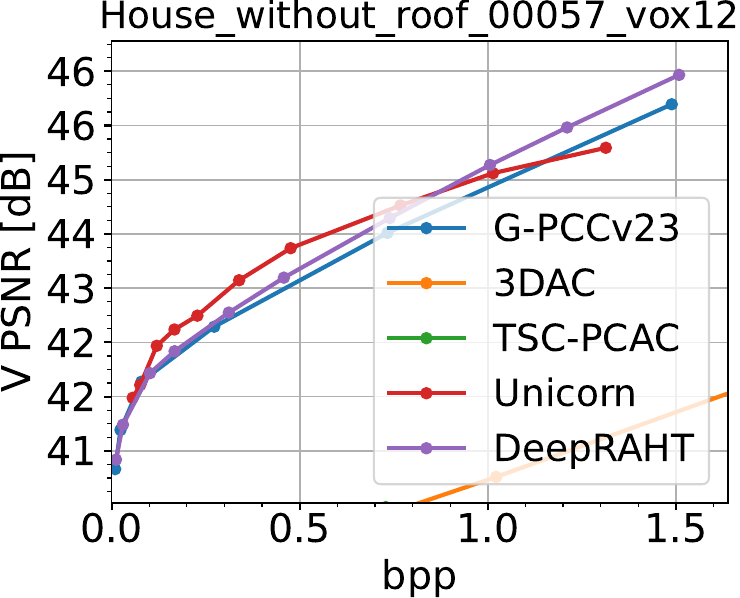}
                }
        
                \subfigure{
                    \includegraphics[width=0.28\textwidth]{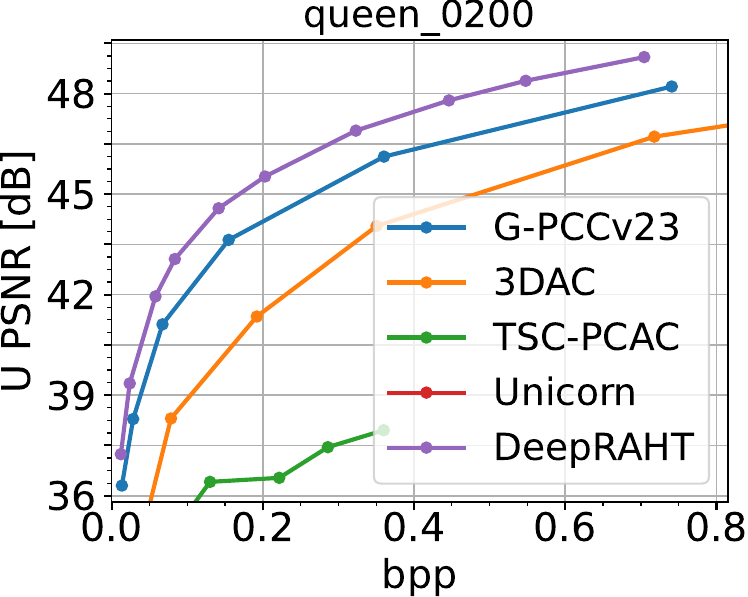}
                }
                \hfill
                \subfigure{
                    \includegraphics[width=0.28\textwidth]{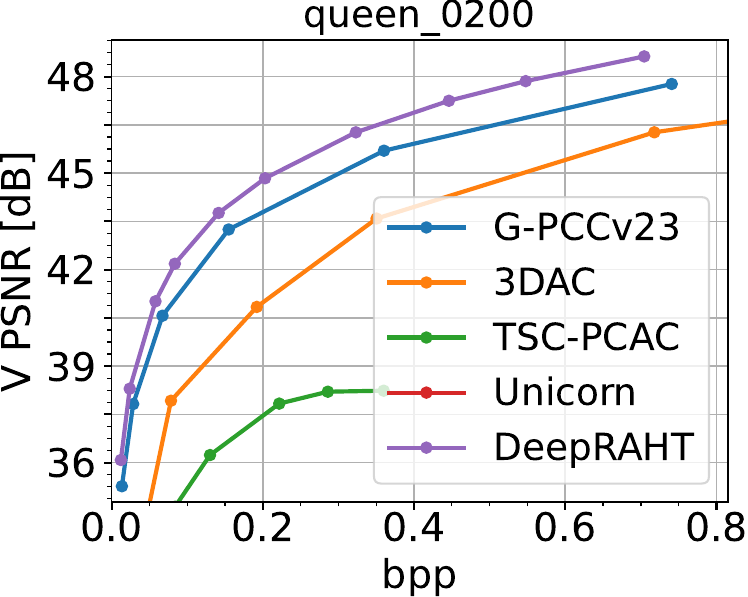}
                }
            \end{adjustbox}
            \begin{adjustbox}{max width=\textwidth} 
        
                \subfigure{
                    \includegraphics[width=0.28\textwidth]{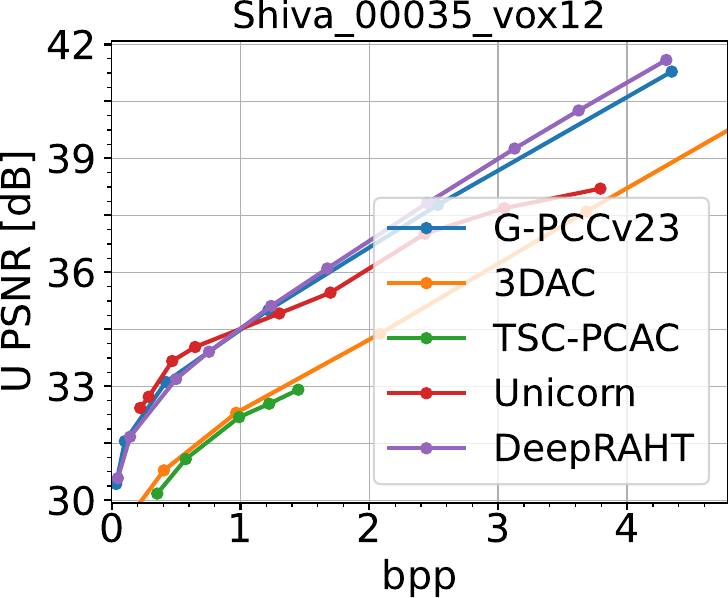}
                }
                \hfill
                \subfigure{
                    \includegraphics[width=0.28\textwidth]{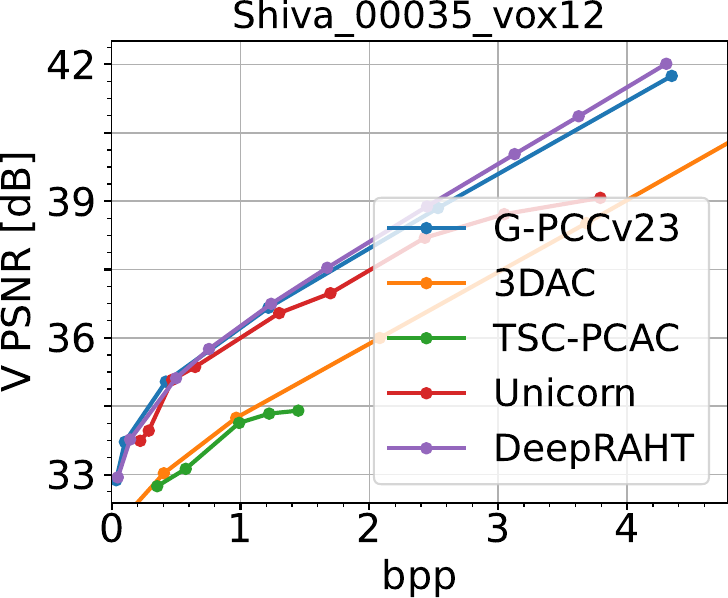}
                }
        
                \subfigure{
                    \includegraphics[width=0.28\textwidth]{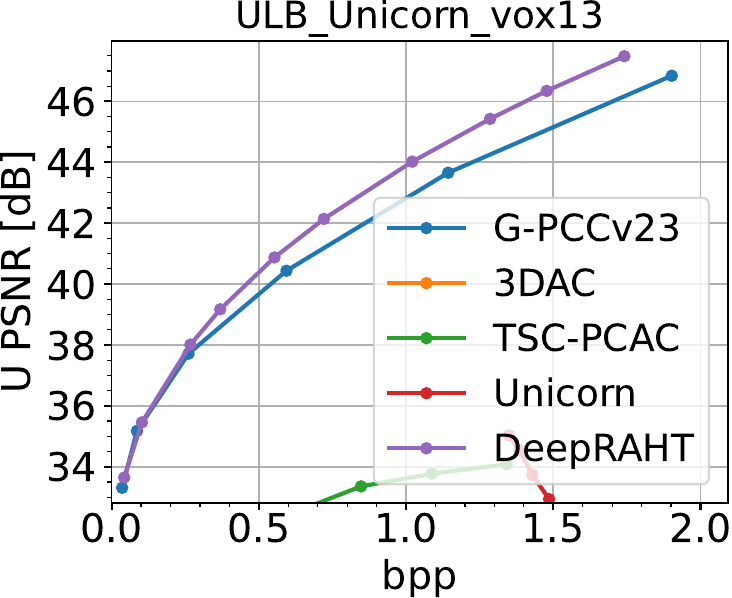}
                }
                \hfill
                \subfigure{
                    \includegraphics[width=0.28\textwidth]{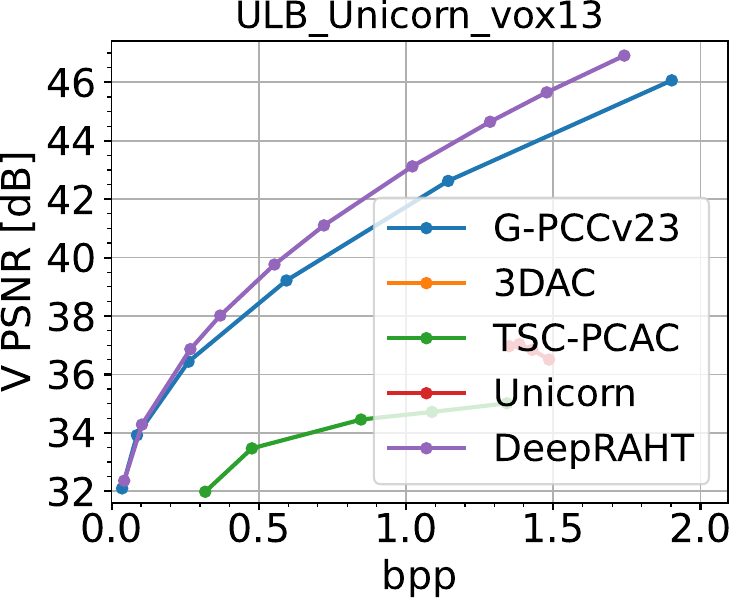}
                }
            \end{adjustbox}
            \begin{adjustbox}{max width=\textwidth} 
    
            \subfigure{
                \includegraphics[width=0.28\textwidth]{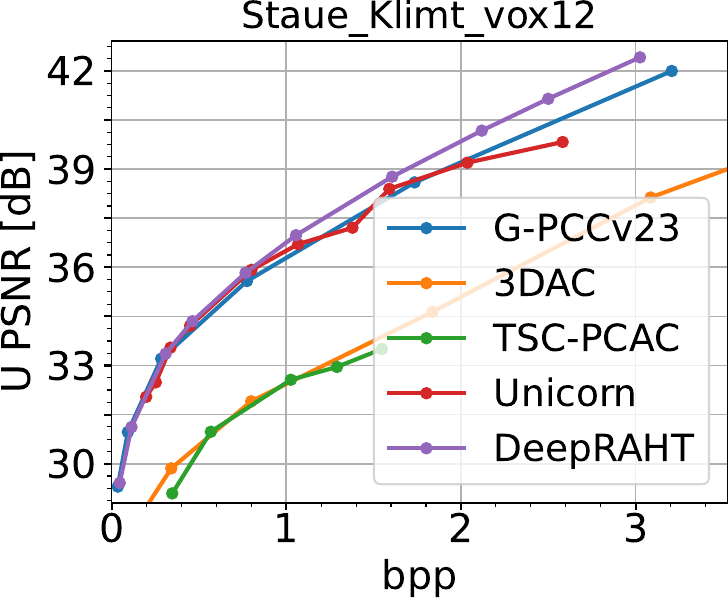}
            }
            \hfill
            \subfigure{
                \includegraphics[width=0.28\textwidth]{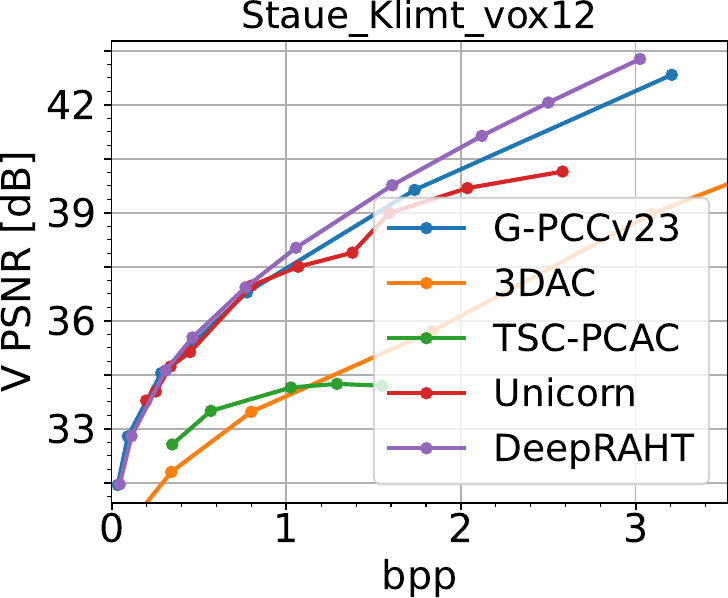}
            }
    
            \subfigure{
                \textcolor{white}{\rule{0.28\textwidth}{0.2\textwidth}} 
            }
            \hfill
            \subfigure{
                \textcolor{white}{\rule{0.28\textwidth}{0.2\textwidth}} 
            }
            \end{adjustbox}
            \caption[]{Detailed chroma R-D comparison of MPEG samples.}
        \end{figure*}
\clearpage
\section{More Qualitative Visualizations}
\begin{figure}[h!]
    \centering
    \includegraphics[scale=0.19]{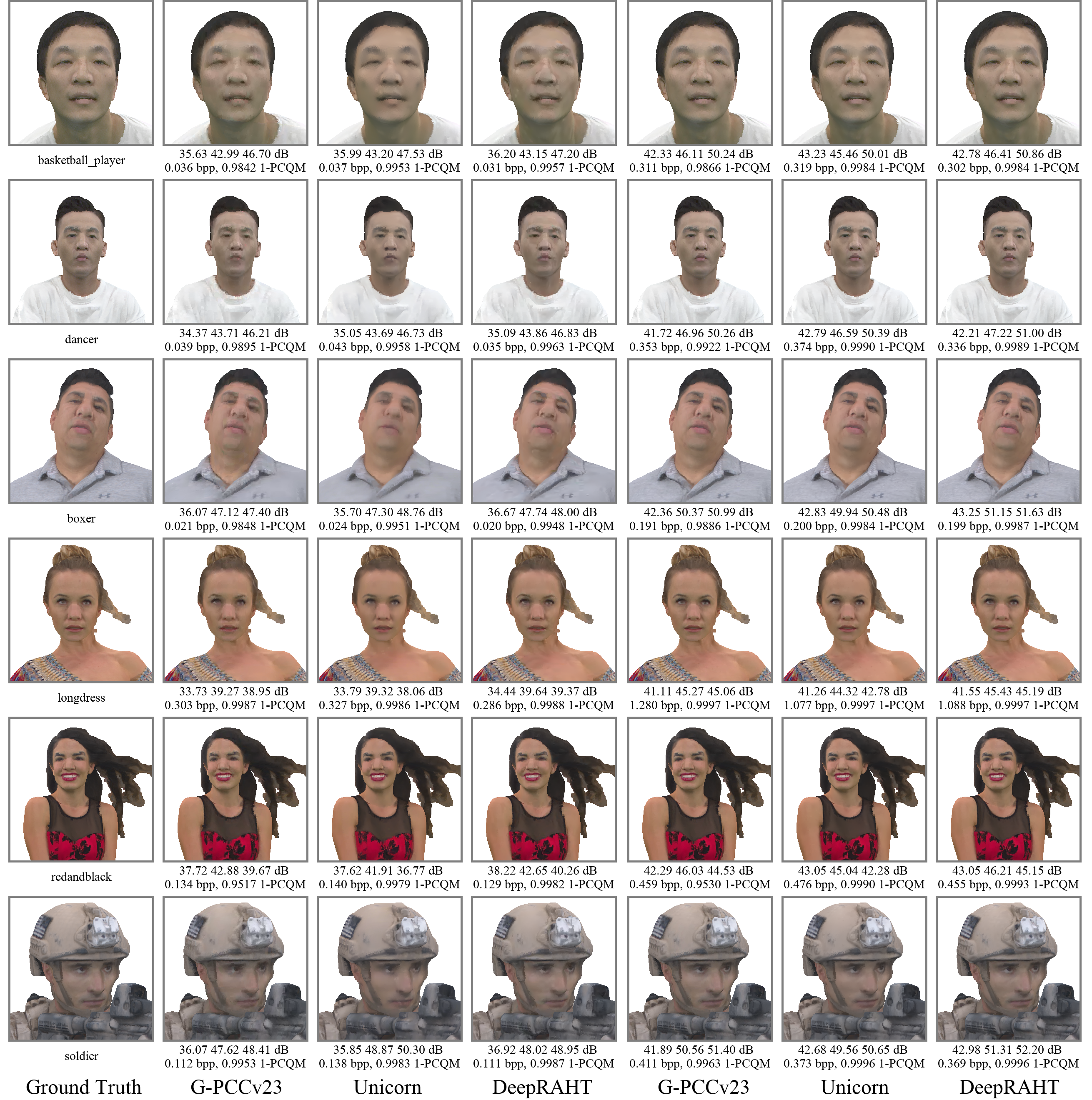}
    \caption{
        More qualitative visualizations. Y, U, V PSNRs, bpp and 1-PCQM are shown under each reconstructed point cloud.
    }
    \label{ab_pred2}
\end{figure}

More qualitative visualizations are shown in Fig.~\ref{ab_pred2}. The left part is the lower quality reconstruction, while the right part is the higher quality reconstruction. PCQM~\cite{meynet2020pcqm} is adopted as the additional quality metric. 

    \bibliography{aaai2026}